\documentclass[useAMS,usenatbib]{mn2e}
\usepackage{color,graphicx,amsmath,ulem}
\def\araa{ARA\&A}       	
\def\apj{ApJ}           	
\def\apjl{ApJ}          	
\def\aap{A\&A}          	
\def\mnras{MNRAS}       	
\def\nat{Nature}        	

\def\gsim{\hspace{0.3em}\raisebox{0.4ex}{$>$}\hspace{-0.75em}\raisebox{-.7ex}{$\sim$}\hspace{0.3em}}
\def\lsim{\hspace{0.3em}\raisebox{0.4ex}{$<$}\hspace{-0.75em}\raisebox{-.7ex}{$\sim$}\hspace{0.3em}}

\title[Spiral-arm instability in MHD]{Spiral-arm instability - II: magnetic destabilisation}
\author[S. Inoue \& N. Yoshida]
{\parbox[t]{\textwidth} 
{Shigeki Inoue\thanks{E-mail: shigeki.inoue@ipmu.jp} \& Naoki Yoshida}
\\ \\
Kavli Institute for the Physics and Mathematics of the Universe (WPI), UTIAS, The University of Tokyo, Chiba 277-8583, Japan\\
Department of Physics, School of Science, The University of Tokyo, Bunkyo, Tokyo 113-0033, Japan
}

\begin{document}

\pagerange{\pageref{firstpage}--\pageref{lastpage}} \pubyear{2014}

\maketitle

\label{firstpage}

\begin{abstract}
Fragmentation of spiral arms can drive the formation of giant clumps and induce intense star formation in disc galaxies. Based on the spiral-arm instability analysis of our Paper I, we present linear perturbation theory of dynamical instability of self-gravitating spiral arms of magnetised gas, focusing on the effect of toroidal magnetic fields. Spiral arms can be destabilised by the toroidal fields which cancel Coriolis force, i.e. magneto-Jeans instability. Our analysis can be applied to multi-component systems that consist of gas and stars. To test our analysis, we perform ideal magneto-hydrodynamics simulations of isolated disc galaxies and examine the simulation results. We find that our analysis can characterise dynamical instability leading arms to fragment and form clumps if magnetic fields are nearly toroidal. We propose that dimensionless growth rate of the most unstable perturbation, which is computed from our analysis, can be used to predict fragmentation of spiral arms within an orbital time-scale. Our analysis is applicable as long as magnetic fields are nearly toroidal. Using our analytic model, we estimate a typical mass of clumps forming from spiral-arm fragmentation to be consistent with observed giant clumps $\sim10^{7-8}~{\rm M_\odot}$. Furthermore, we find that, although the magnetic destabilisation can cause low-density spiral arms to fragment, the estimated mass of resultant clumps is almost independent from strength of magnetic fields since marginal instability occurs at long wavelengths which compensate the low densities of magnetically destabilised arms.

\end{abstract}


\section{Introduction}
\label{Intro}
Dynamical instability in disc galaxies can play an important role in their formation and evolution processes. In the high-redshift Universe, a significant fraction of galaxies have been observed to have clumpy morphologies in which several star-forming `giant clumps' of $M_{\rm cl}\lsim10^8~{\rm M_\odot}$ are hosted in their disc regions; $\sim50$ per cent of galaxies of $M_{\rm star}\simeq10^9$--$10^{11}~{\rm M_\odot}$ at a redshift of $z\simeq1.5$--$3$ are observed to be clumpy \citep{tkt:13II,mkt:14,gfb:14,sok:16}.\footnote{The abundance of clumpy galaxies is observed to depend on galaxy mass and redshift; it decreases for massive galaxies in low-redshift Universe.} Although their clumpy morphologies can be attributed to ongoing mergers in some galaxies \citep[e.g.,][]{w:06,fgb:09,p:10,r:17}, the clumpy galaxies are observed to have significant rotations indicative of their disc structures \citep[e.g.][]{g:06,g:08,gnj:11}. 

It is has been discussed that the high-redshift clumpy galaxies may evolve to spiral galaxies in the present-day Universe \citep[e.g.][]{SN:93,n:98,n:99} and that giant clumps may be related to formation processes and dynamical properties of galactic structures such as discs, bulges, dark matter haloes and halo objects \citep[e.g.][]{eev:05,ebe:08,ee:06,bee:07,es:13,is:11,is:12,is:14,nb:15,i:13}. It has also been proposed, on the other hand, that the giant clumps observed are transient structures and are soon disrupted within an orbital time-scale \cite[e.g.][]{ho:10,g:12}. Cosmological simulations of \citet{bmo:17,okh:17} showed that these clumpy structures are bright in star-forming light such as H$\alpha$ but not massive enough to be self-gravitating within galactic potential. Clumpy disc galaxies are also observed at low redshifts although their abundance is lower by far than at high redshifts \citep[e.g.][]{ees:13,bgf:14,fgb:14,f:17,gpm:15}. 
 
The physical mechanism of clump formation is often attributed to local gravitational instability of radial perturbations in galactic discs (e.g. \citealt{n:99,dsc:09,gnj:11,fga:17} although see \citealt{idm:16}), i.e. Toomre instability \citep{s:60,t:64}. In our Paper I \citep[][]{iy:18}, for the first time, we have proposed spiral-arm instability (SAI) model for the giant clump formation, in which we consider local fragmentation of spiral arms, rather than discs, against azimuthal perturbations. In our Paper I, we argue that our SAI model describes dynamical properties of giant clumps observed in low-redshift galaxies (see section 5 of Paper I). Thus, fragmentation of spiral arms could be a possible mechanism of giant clump formation. Our  linear perturbation analysis for spiral arms presented in our Paper I can characterise fragmentation of spiral arms quite accurately even in two-component models of gas and stars. This means that spiral-arm fragmentation is basically considered to be a linear process that can be described as balance between self-gravity, pressure and Coriolis force.

In this paper, we study instability of spiral arms in the context of ideal magneto-hydro dynamics (MHD) by taking into account effects of toroidal magnetic field in our SAI model. In previous studies, magnetic effects are incorporated in linear perturbation analysis for a self-gravitating local region in a rotating flat disc \citep[e.g.][]{l:66,e:87,e:94,g:96,ko:01}. In these analyses, spiral arms are considered as perturbations propagating on a disc. \citet{e:94} discusses gravitational instability in the crest of a spiral density wave as swinging azimuthal perturbations in a disc. \citet{ko:01} perform ideal MHD simulations with their shearing-box model and show that spiral arms can be destabilised and fragment into clumps when azimuthal magnetic fields are strong. Thus, as was shown in the previous works, toroidal magnetic fields can induce fragmentation of spiral arms and may drive formation of giant clumps. Unlike Toomre analysis for disc instability, our SAI analysis is adopted to geometry of a local ring structure (a tightly would spiral arm), therefore thought to be better at describing fragmentation of spiral arms. We expect that our SAI analysis including the magnetic effect could give us more accurate description of spiral-arm fragmentation induced by azimuthal magnetic fields.

This paper is organised as follows. In Section \ref{basiceq}, based on our Paper I, we present our linear perturbation analysis including the effects of magnetic fields and characterisation of the instability of perturbations. Furthermore, we develop our theory to a two-component model that consists of gas and stars. In Section \ref{sims}, we perform ideal MHD simulations with isolated disc galaxy models to test our theory. In Section \ref{result}, we adopt our simulation data to our instability analysis and examine our linear analysis. In Section \ref{discussion}, we discuss how masses of giant clumps forming via SAI are affected by strength of magnetic fields and estimate a typical clump mass. In addition, we argue more general effects of magnetic fields on dynamics of galactic discs and whether disc galaxies and their spiral arms are destabilised by magnetic fields in reality. In Section \ref{conclusions}, we present our conclusions and summary of this study.

\section{Linear perturbation analysis of ideal MHD}
\label{basiceq}
Our analysis presented in Paper I basically follows that proposed by \citet{tti:16} for a single-component non-magnetised gas disc. The local linear perturbation analysis assumes that a pitch angle of a spiral arm to be negligibly small: the tight-winding approximation, where the spiral arm can be locally approximated as a structure resembling a ring. The arm is assumed to have rigid rotation with an angular velocity $\Omega$ since the arm is expected to be self-gravitating (see Appendix B of Paper I); therefore the Oort's constant $B=-\Omega$ in the arm. In the polar coordinates $(R,\phi)$, we consider azimuthal perturbations propagating inside the arm, which are assumed to be proportional to $\exp[\mathrm{i}(ky-\omega t)]$, where $y\equiv\phi R$. When $\omega$ has a positive imaginary part, the perturbation is expected to grow exponentially with time and thus dynamically unstable. For the perturbations, if their wavelengths are small enough compared with the radius of the arm, i.e. $kR\gg1$, then the curvature of the spiral arm is negligible. 

We adopt a Gaussian distribution to a radial surface density profile of the spiral arm, $\Sigma(R)=\Sigma_0\exp(-\xi^2/2w^2)$, where $\xi\equiv R-R_0$, $R_0$ is the radius of the density peak in the arm, and $\Sigma_0$ is the surface density at $R_0$. As is done in \citet{tti:16} and our Paper I, we define the edges of the spiral arm to be the inner and outer radii at which $\Sigma(R)=0.3\Sigma_0$. In this case, the half arm width is $W\simeq1.55w$. The line-mass of the arm is given as 
\begin{equation}
\Upsilon\equiv2\int^W_0\Sigma(R)~\textrm{d}\xi=AW\Sigma_0,
\label{linemass}
\end{equation}
where $A\simeq1.4$ for a Gaussian density distribution.

In this Section, we present our linear perturbation analysis for spiral arm taking into account the effect of azimuthal magnetic field. Our models for single- and two-component spiral arms are considered in Section \ref{ana_single} and \ref{2compana}. 

\subsection{Single-component analysis}
\label{ana_single}
We consider a magnetized gas disc. In a reference frame rotating with the disc at an angular velocity $\Omega$, the equations of continuity, momentum and magnetic conservation are given as 
\label{eqs}
\begin{equation}
\frac{\partial \rho}{\partial t}+\nabla\cdot\left(\rho\mathbfit{v}\right)=0,
\label{Continuity}
\end{equation}
\begin{equation}
\frac{\partial \mathbfit{v}}{\partial t}+\left(\mathbfit{v}\cdot\nabla\right)\mathbfit{v}+2\bmath{\Omega}\times\mathbfit{v}-\Omega^2R=-\frac{\nabla p}{\rho}-\nabla\Phi-\frac{\mathbfit{B}\times\left(\nabla\times\mathbfit{B}\right)}{4\pi\rho},
\label{Momentum}
\end{equation}
\begin{equation}
\frac{\partial \mathbfit{B}}{\partial t} = \nabla\times\left(\mathbfit{v}\times\mathbfit{B}\right).
\label{Faraday}
\end{equation}
In equation (\ref{Faraday}), we do not take into account magnetic diffusion effects, i.e. ideal MHD. In our analysis, we ignore vertical structures of spiral arms and assume the magnetic pressure is uniform within the spiral arm. 

In observations, magnetic fields in nearby spiral galaxies are approximately oriented along their spiral arms with the mean strength of a few ${\rm \mu G}$ to about $20~{\rm \mu G}$ \citep[][and references therein]{h:17}. Since our analysis assumes spiral arm to be tightly wound, we consider the equilibrium magnetic field $\mathbfit{B}_0$ is parallel to the azimuthal direction. Then, perturbed magnetic field is written as $\mathbfit{B}=\{\delta B_R,B_0+\delta B_\phi, 0\}$. By considering two-dimensional space, spatial density $\rho$ is replaced with surface density $\Sigma$. Using the aforementioned assumptions and settings, the linearlised equations of continuity, $R$- and $\phi$-momenta for the azimuthal perturbations are obtained as
\begin{equation}
\omega\delta \Upsilon = k\Upsilon\delta v_\phi,
\label{linearlized1}
\end{equation}
\begin{equation}
-\mathrm{i}\omega\delta v_R = 2\Omega\delta v_\phi - \mathrm{i}\frac{k^2}{\omega}v_{\rm A}^2\delta v_R,
\label{linearlized2}
\end{equation}
\begin{equation}
-\mathrm{i}\omega\delta v_\phi = -2\Omega\delta v_R - \mathrm{i}k\frac{\sigma^2}{\Upsilon}\delta \Upsilon - \mathrm{i}k\delta\Phi
\label{phimom}
\end{equation}
where the prefix $\delta$ means the perturbation of a physical value following it, and Alfv\'{e}n velocity $v_{\rm A}^2\equiv B_{0}^2/(4\pi\rho)$.\footnote{Assuming the disc to be vertically uniform, vertically-integrated magnetic pressure is approximated as $B_{\rm 2D}^2/2=B_{\rm 3D}^2h/2$, where disc thickness $h\equiv\Sigma/\rho$, and $B_{\rm 2D}$ and $B_{\rm 3D}$ are magnetic fields in two- and three-dimensional space. In the two-dimensional space, therefore, vertically-integrated Alfv\'{e}n velocity $v_{\rm A,2D}^2\equiv B_{\rm 2D}^2/(4\pi\Sigma)=B_{\rm 3D}^2/(4\pi\rho)$; therefore $v_{\rm A,2D}=v_{\rm A,3D}$ \citep[see][]{ko:02}.} For the pressure term in equation (\ref{phimom}), a barotropic equation of state is assumed, and $\sigma^2\equiv\sigma_\phi^2+c_{\rm snd}^2$, where $\sigma_\phi$ and $c_{\rm snd}$ are azimuthal dispersion of turbulent velocities and sound velocity of gas. 

The Poisson equation for the perturbed potential of a razor-thin ring with a Gaussian density distribution is given as
\begin{equation}
\label{poisson}
  \delta\Phi =-\pi G\delta\Upsilon\left[K_0(kW)L_{-1}(kW) + K_1(kW)L_0(kW)\right],
\end{equation}
where $K_i$ and $L_i$ are modified Bessel and Struve functions of order $i$ \citep{tti:16}. Hereafter, we denote $f(kW)\equiv[K_0(kW)L_{-1}(kW) + K_1(kW)L_0(kW)]$. The function $f(kW)$ decreases with $kW$. For $kW\gg1$ (i.e. short wavelength $\lambda\ll W$), approximately $f(kW)\propto(kW)^{-1}$, therefore the Poisson equation becomes similar to that of a uniform disc. For $kW\ll1$ (i.e. long wavelength $\lambda\gg W$), the slope of $f(kW)$ becomes shallower, therefore the Poisson equation deviates from that used in Toomre analysis (see fig. 1 of our Paper I).

Combining equations from (\ref{linearlized1}) to (\ref{poisson}), one can obtain a dispersion relation for azimuthal perturbations within a magnetized spiral arm, 
\begin{equation}
\omega^2 = \left[\sigma^2-\pi G\Upsilon f(kW)\right]k^2 + \frac{4\Omega^2\omega^2}{\omega^2-k^2v_{\rm A}^2}.
\label{DR}
\end{equation}
Similar analyses deriving a dispersion relation for magnetised disc have been presented in previous studies \citep[e.g.][]{l:66,e:87,e:94,g:96,ko:01}. The last term in the right-hand side represents the contribution of Coriolis force which stabilises large-scale perturbations with small $k$. In order to understand the effect of toroidal fields, we consider an unstable state with $\omega^2<0$, and replace $\omega^2$ with $-|\omega^2|$ in the last term. In this case, the dispersion relation (\ref{DR}) becomes
\begin{equation}
\omega^2 = \left[\sigma^2-\pi G\Upsilon f(kW)\right]k^2 + \frac{4\Omega^2|\omega^2|}{|\omega^2|+k^2v_{\rm A}^2}.
\end{equation}
As can be seen in the above equation, toroidal magnetic fields cancel the Coriolis force and can destabilise large-scales perturbations in the arm. The effect canceling Coriolis force can be intuitively understood as follows. Azimuthal perturbations drive velocity perturbations $\delta v_\phi$ along a spiral arm. Because both $\delta v_\phi$ and the magnetic field are azimuthal, the perturbations are not directly affected by the field. However, Coriolis force exerts perpendicular to $\delta v_\phi$, i.e. the Coriolis force is radial. Since the magnetic field is azimuthal, the magnetic force exerts radial and thus counter-acts the Coriolis force.

Toomre's instability parameter measured in a spiral arm with rigid rotation is given as 
\begin{equation}
Q_{\rm sp}\equiv \frac{2\sigma\Omega}{\pi G\Sigma}=\frac{2A\sigma\Omega W}{\pi G\Upsilon}.
\label{ToomreQ}
\end{equation}
Influence on gravitational potential by finite thickness of the spiral arm can be taken into account in $Q_{\rm sp}$ although we assume an  infinitesimally thin density distribution for the spiral arm.\footnote{Previous studies have proposed models of the thickness correction for disc potential \citep[e.g.][]{gl:65,r:92,e:11,rw:11,bbs:14}.} To make the dispersion relation dimensionless, we introduce normalised frequency and wavelength of perturbations as $s\equiv\omega/(2\Omega)$ and $x\equiv kW$. With $q\equiv \sigma/(2\Omega W)$ and $\beta\equiv \sigma^2/v_{\rm A}^2$,\footnote{This quantity is similar to plasma-$\beta$ but different in terms of taking into account turbulent pressure.} the dispersion relation (\ref{DR}) is written as 
\begin{equation}
s^2 = \left[q^2 - \frac{Aq}{Q_{\rm sp}}f(x)\right]x^2 + \frac{s^2}{s^2-\beta^{-1}q^2x^2}.
\label{ND_DR1}
\end{equation}
In addition, we denote
\begin{equation}
J(x)\equiv q^2 - \frac{Aq}{Q_{\rm sp}}f(x).
\label{Jeans}
\end{equation}
Then, equation (\ref{ND_DR1}) is reduced to 
\begin{equation}
s^4 - \left[\beta^{-1}q^2x^2 + J(x)x^2 + 1\right]s^2 + J(x)x^4\beta^{-1}q^2 = 0.
\label{ND_DR2}
\end{equation}
Because this dispersion relation is a biquadratic equation of $s$, the two roots are obtained by the quadratic formula as
\begin{equation}
\begin{split}
s^2 &= \frac{\beta^{-1}q^2x^2 + J(x)x^2 + 1}{2} \\
&\quad \pm \frac{\sqrt{\left[\beta^{-1}q^2x^2 + J(x)x^2 + 1\right]^2 - 4J(x)x^4\beta^{-1}q^2}}{2}.
\end{split}
\label{quadra}
\end{equation}
Let $s_+^2$ and $s_-^2$ denote the roots that take the positive and negative signs for the second term, respectively.  Because $s_+^2>0$ for all $x$, this root does not indicate instability. It is unimportant for our SAI analysis. On the other hand, $s_-^2<0$ when $J(x)<0$, and it corresponds to unstable states. This instability condition, $J(x)<0$, reduces to
\begin{equation}
\sigma^2-\pi G\Upsilon f(kW)<0,
\end{equation}
which is equivalent to the Jeans instability condition in absence of disc rotation and magnetic effect. Gravity dominates over thermal and turbulent pressure when $J(x)<0$. Thus, the solution of $s_-^2<0$ corresponds to `magneto-Jeans instability' \citep[][]{ko:01,ko:02}, in which toroidal magnetic fields cancel the stabilising effect of Coriolis force, and then perturbations can collapse unless pressure is strong enough to resist gravity. Large-scale perturbations with small $x$ necessarily satisfy the condition $J(x)<0$ in all range below a certain $x$ that gives $J(x)=0$. In the presence of toroidal magnetic fields (i.e. $v_{\rm A}\ne0$), spiral arms are thus unstable in large scales: magnetic destabilisation.

We define the dimensionless growth rate $s_{\rm grow}$ for a perturbation from the root $s_-^2$ of equation (\ref{quadra}). For stable perturbations with $s_-^2>0$, the growth rates are defined to be $s_{\rm grow}(x)=0$, whereas for unstable perturbations with $s_-^2<0$, $s_{\rm grow}(x)=\sqrt{-s_-^2}$. In this study, since we focus on the perturbation that grows and collapses first, we consider $\max[s_{\rm grow}(x)]$ to be a quantity that describes the degree of instability in a local region.

\begin{figure}
	\includegraphics[bb=0 0 887 831, width=\hsize]{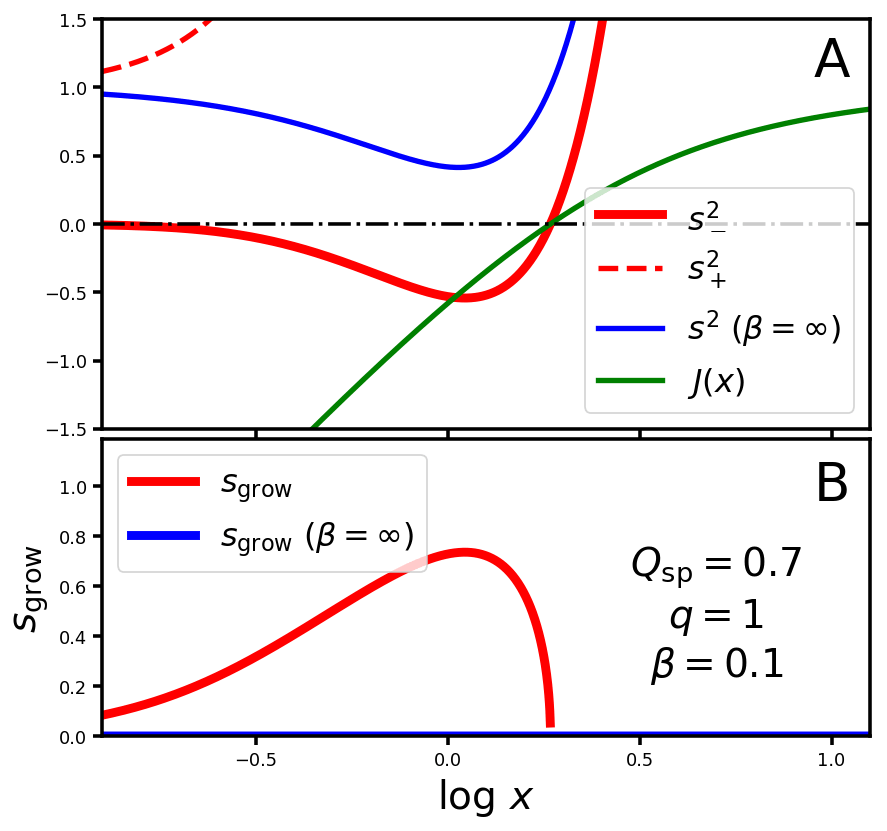}
	\includegraphics[bb=0 0 887 831, width=\hsize]{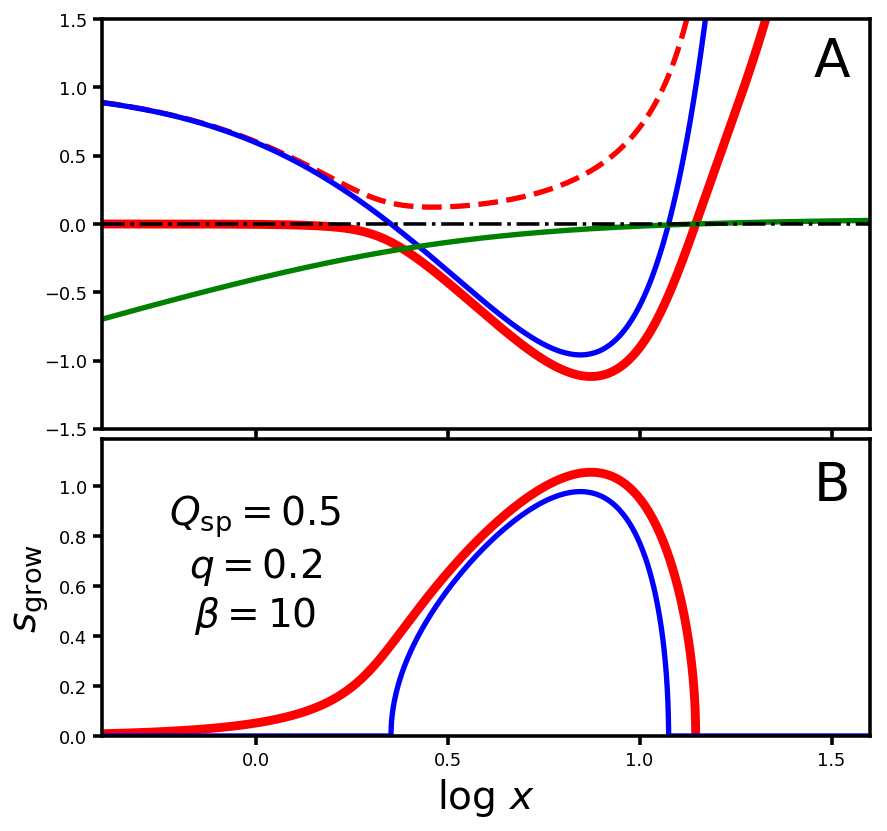}
	\caption{Solutions of the dispersion relations $s_-^2\equiv[\omega/(2\Omega)]^2$ (red solid lines in Panels A) and dimensionless growth rates $s_{\rm grow}$ (red lines in Panels B) in the single-component analysis, as functions of dimensionless wavenumber $x\equiv kW$. These are computed by solving equation (\ref{quadra}). The dotted red lines in Panels A correspond to the counterparts $s_+^2$ that take the positive sign in equation (\ref{quadra}). The blue lines indicate $s^2$ and $s_{\rm grow}$ for non-magnetic solutions with $v_{\rm A}=0$ in Panels A and B. The green lines in Panels A indicate the Jeans instability condition $J(x)$; the unstable states with $s_-^2<0$ appear when $J(x)<0$. The top and bottom sets of panels show results with different parameter sets for $Q_{\rm sp}$, $q$ and $\beta$.}
	\label{DRfigs}
\end{figure}
As seen in equation (\ref{ND_DR1}), the dimensionless dispersion relation $s^2$ as a function of $x$ is described with the three parameters: $Q_{\rm sp}$, $q$ and $\beta$. Fig. \ref{DRfigs} shows the dispersion relations for two example cases. The top set of panels shows the case where $Q_{\rm sp}=0.7$, $q=1$ and $\beta=0.1$. In Panel A, we compare the solutions of $s_-^2$, $s_+^2$ and that in non-magnetic case ($\beta=\infty$). Panel B shows $s_{\rm grow}(x)$ for the magnetic (red) and non-magnetic (blue) cases. Note that $s_{\rm grow}=0$ for all $x$ in the non-magnetic case since all perturbations are stable with this parameter set. Large-scale perturbations with small $x$ are necessarily unstable, and indicate $s_-^2<0$ and $s_{\rm grow}>0$ as long as $v_{\rm A}\ne0$. However, $s_-^2$ and $s_{\rm grow}$ approach asymptotically to $0$ as $x$ decreases. Such large-scale perturbations with $s_{\rm grow}\simeq0$ can be considered to be practically stable. Hence, we can regard the perturbation $x$ that gives $\max(s_{\rm grow})$ as the characteristic perturbation of the instability. The bottom set of panels in Fig. \ref{DRfigs} shows the same result but for the case where $Q_{\rm sp}=0.5$, $q=0.2$ and $\beta=10$. With this parameter set, perturbations are unstable even if there are no magnetic fields. The values of $s_-^2$ are smaller than $s^2$ with $\beta=\infty$ in all $x$, and this means that the toroidal magnetic fields enhance the gravitational instability.

\begin{figure}
  \includegraphics[bb=0 0 2343 2715, width=\hsize]{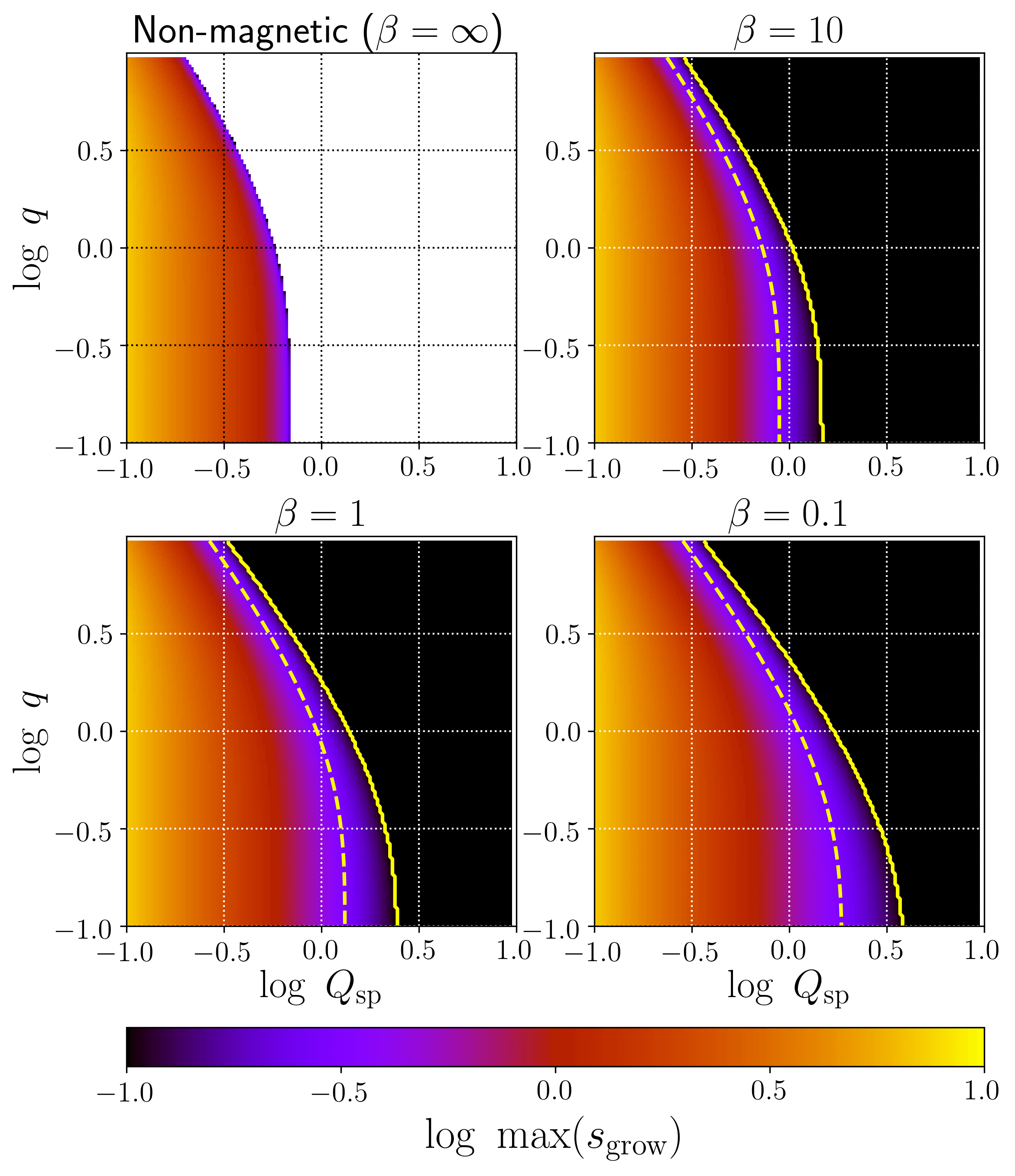}
  \caption{The maximum growth rates, $\max[s_{\rm grow}(x)]$ computed with the single-component analysis, as a function of $Q_{\rm sp}\equiv2A\sigma\Omega W/(\pi G\Upsilon)$ and $q\equiv \sigma/(2\Omega W)$. The top left panel is for the non-magnetic case with $v_{\rm A}=0$, and the other panels show the results with $\beta\equiv\sigma^2/v_{\rm A}^2=10$ (top right), $1$ (bottom left) and $0.1$ (bottom right). The coloured regions are unstable ($\max[s_{\rm grow}(x)]>0$), and the uncoloured regions in the top left panel are stable ($s_{\rm grow}=0$ for all $x\equiv kW$). Weakly unstable regions with $\max[s_{\rm grow}(x)]<0.1$ are coloured black. The yellow solid and dashed lines delineate the contours of $\max(s_{\rm grow})=0.1$ and $0.3$; these contours are not shown in the top left panel since they almost overlap with the boundary between the stable and unstable regions.}
  \label{DRall}
\end{figure}	
Fig. \ref{DRall} illustrates $\max[s_{\rm grow}(x)]$ as functions of $Q_{\rm sp}$ and $q$ for $\beta=\infty$, $10$, $1$ and $0.1$. The top left panel shows the result of the non-magnetic case, in which uncoloured regions indicate stable state ($s_{\rm grow}=0$ for all $x\equiv kW$), and we can definitely find the boundary between the stable and unstable states in this case. On the other hand, we cannot specify exact boundaries in magnetic cases since large-scale perturbations are unstable and $\max(s_{\rm grow})>0$ in the presence of toroidal fields; however the regions with $\max[s_{\rm grow}(x)]\ll1$ are expected to be practically stable. It can be seen that regions with large $\max(s_{\rm grow})$ spread to high $Q_{\rm sp}$ with decreasing $\beta$ for strong magnetic fields (see the yellow lines). The destabilising effect is more significant for smaller $q$ at a given $Q_{\rm sp}$.

In disc galaxies, emergence of spiral-arm structures is thought to start with low $\Sigma$ and $v_{\rm A}$, where perturbations are expected to be stable with $\max(s_{\rm grow})\ll1$. The arms grow gradually and eventually reach practically unstable states with somewhat large $\max(s_{\rm grow})$, where perturbations can start growing. Therefore, the onset of instability is considered to occur at the weakly unstable states where $\max(s_{\rm grow})\sim0.3$ (see Section \ref{Criterion}). In Section \ref{ClumpMassDiscuss}, we discuss a typical mass of clumps that form from this marginal instability.

\subsection{Two-component analysis}
\label{2compana}
Next, we extend the above linear analysis to a multi-component model composed of magnetised gas and stars. We consider distinct values of $\Upsilon$, $\sigma$ and $W$ for each of the gas and stellar component. Hereafter, let suffixes `g' and `s' denote gas and stellar values in the two-component analysis. In this study, we assume that angular rotation velocities are the same between the gas and stars, i.e. $\Omega\equiv\Omega_{\rm g}=\Omega_{\rm s}$.\footnote{Although the analysis presented in our Paper I does not assume $\Omega_{\rm g}=\Omega_{\rm s}$, only small differences of $\overline{v_{\rm \phi}}$ between gas and stars are found in isolated simulations of Paper I and cosmological simulations of \citet{idm:16}.} We adopt the fluid approximation to the stellar component, in which stars are assumed to have the same form of dispersion relation of gas. Although the fluid approximation for stars may not be appropriate because of the collisionless nature of stars \citep{t:64,ls:66,bt:08,r:01,e:11}, our Paper I has demonstrated that the fluid approximation does not deteriorate accuracy of our SAI analysis and that our linear analysis can characterise fragmentation of arms in $N$-body disc simulations.

For the gas component, by combining equations (\ref{linearlized1}, \ref{linearlized2} and \ref{phimom}), a perturbed line-mass of gas is obtained as 
\begin{equation}
\delta \Upsilon_{\rm g} = k^2\frac{\Upsilon_{\rm g}}{\omega^2-\frac{4\Omega^2\omega^2}{\omega^2-k^2v_{\rm A}^2}-\sigma_{\rm g}^2k^2}\delta\Phi.
\label{DRgas}
\end{equation}
On the other hand, stars are not affected by magnetic fields, therefore 
\begin{equation}
\delta \Upsilon_{\rm s} = k^2\frac{\Upsilon_{\rm s}}{\omega^2-4\Omega^2-\sigma_{\rm s}^2k^2}\delta\Phi,
\label{DRstar}
\end{equation}
where $\sigma_{\rm s}$ is azimuthal component of stellar velocity dispersion. The gas and stars share the same perturbed potential that is described as $\delta\Phi=\delta\Phi_{\rm g}+\delta\Phi_{\rm s}$ \citep{js:84_2,js:84,r:92,j:96,r:01}, and the Poisson equation (\ref{poisson}) connects the two components as
\begin{equation}
\label{twopoisson}
  \delta\Phi =-\pi G\left[\delta \Upsilon_{\rm g}f(kW_{\rm g}) + \delta \Upsilon_{\rm s}f(kW_{\rm s})\right].
\end{equation}
Thus, the two-component dispersion relation is obtained as 
\begin{equation}
\frac{\pi Gk^2\Upsilon_{\rm g}f(kW_{\rm g})}{\sigma_{\rm g}^2k^2+\frac{4\Omega^2\omega^2}{\omega^2-k^2v_{\rm A}^2}-\omega^2} + \frac{\pi Gk^2\Upsilon_{\rm s}f(kW_{\rm s})}{\sigma_{\rm s}^2k^2+4\Omega^2-\omega^2}= 1.
\label{DR2C}
\end{equation}
As for equation (\ref{ND_DR1}), we introduce dimensionless quantities for each of gas and stars, then we obtain
\begin{equation}
\frac{AQ_{\rm sp,g}^{-1}q_{\rm g}f(x_{\rm g})x_{\rm g}^2}{q_{\rm g}^2x_{\rm g}^2+\frac{s^2}{s^2-\beta^{-1}q_{\rm g}^2x_{\rm g}^2}-s^2}  + \frac{AQ_{\rm sp,s}^{-1}q_{\rm s}f(x_{\rm s})x_{\rm s}^2}{q_{\rm s}^2x_{\rm s}^2+1-s^2}= 1.
\label{DR2Cnd}
\end{equation}

Furthermore, by denoting $\gamma_{\rm g}\equiv AQ_{\rm sp,g}^{-1}f(x_{\rm g})q_{\rm g}x_{\rm g}^2$, $\gamma_{\rm s}\equiv AQ_{\rm sp,s}^{-1}f(x_{\rm s})q_{\rm s}x_{\rm s}^2$, $\alpha_{\rm g}\equiv q_{\rm g}^2x_{\rm g}^2-\gamma_{\rm g}$ and $\alpha_{\rm s}\equiv q_{\rm s}^2x_{\rm s}^2+1-\gamma_{\rm s}$, the dispersion relation (\ref{DR2Cnd}) can be reduced to a bicubic equation of $s$:
\begin{equation}
\begin{split}
&s^6 -\left(\beta^{-1}q_{\rm g}^2x_{\rm g}^2 + \alpha_{\rm s} + \alpha_{\rm g} + 1\right)s^4 \\
&+\left[\alpha_{\rm s}\alpha_{\rm g} + \beta^{-1}q_{\rm g}^2x_{\rm g}^2\left(\alpha_{\rm s}+\alpha_{\rm g}\right) +\alpha_{\rm s}-\gamma_{\rm s}\gamma_{\rm g}\right]s^2\\
&+\beta^{-1}q_{\rm g}^2x_{\rm g}^2\left(\gamma_{\rm s}\gamma_{\rm g} - \alpha_{\rm s}\alpha_{\rm g}\right) = 0.
\end{split}
\label{sixthorder}
\end{equation}
Since the algebraic solutions of a cubic equation can be derived in a number of different ways such as Cardano's and Vieta's methods \citep[e.g.][]{w:91}, one can compute the three roots of the bicubic equation (\ref{sixthorder}) for each $k$. All of the three roots $s^2$ are always real numbers. Two of them are positive for all $k$, do not correspond to unstable modes. Only the smallest root of the three can be negative, and we denote this root as $s_-^2$ and compute the dimensionless growth rate $s_{\rm grow}$ from $s_-^2$. The values of $s_-^2$ and $s_{\rm grow}$ behave similar to those in the single-component case shown in Figs. \ref{DRfigs} and \ref{DRall}. Namely, although $s_-^2$ necessarily becomes negative for small $k$ and indicates instability for large-scale perturbations, it approaches asymptotically to $s_-^2=0$ as $k$ decreases, and $s_{\rm grow}$ has a maximum at a characteristic $k$. If $\max[s_{\rm grow}(k)]\ll1$, the local region is only subject to weak instability, and is essentially stable.

\section{Simulations}
\label{sims}
To test our theory proposed in Section \ref{basiceq}, we perform numerical simulations of isolated disc galaxies. We use the moving-mesh MHD/$N$-body code {\sc Arepo} \citep{arepo} to perform simulations of self-gravitating discs that consist of stars and magnetised gas with an isothermal equation of state. The effect of magnetic fields are assumed to be ideal MHD and implemented with the Powell approach \citep{prl:99} for keeping $\nabla\cdot\mathbfit{B}$ negligibly small \citep{ps:13}. Our simulations do not take into account gas cooling, star formation, stellar feedback or magnetic diffusion.

\subsection{Initial conditions}
Our initial conditions of the galactic discs are similar to those used in our Paper I, which are generated with the method proposed by \citet{h:93}. The initial profile of radial velocity dispersions are assumed to follow that of surface densities described with an exponential function with the scale radius $R_{\rm d}=3~{\rm kpc}$, and kinematic coldness is parameterised by a Toomre parameter $Q_{\rm min}$ at $R\simeq2.5R_{\rm d}$.\footnote{In computing $Q_{\rm min}$ in the initial conditions, we do not take into account thermal or magnetic pressure but only turbulent velocity dispersion is included.} Vertical structures of the discs are constructed with a density function of ${\rm sech}^2[z/(2z_{\rm d})]$ with a constant scale height $z_{\rm d}=50~{\rm pc}$ and a velocity distribution determined from vertical Jeans equilibrium. 

For the single-component model, the total mass of the gas disc is $M_{\rm d}=1.24\times10^{10}~{\rm M_\odot}$, and we set $Q_{\rm min}=2$. For the two-component model, the total mass of gas and stellar discs are $M_{\rm d,g}=7.22\times10^9~{\rm M_\odot}$ and $M_{\rm d,s}=3.40\times10^{10}~{\rm M_\odot}$, corresponding to the gas fraction $f_{\rm g}=0.175$, with $Q_{\rm min}=1.5$. The gas and stars initially share the same density and velocity distributions described above. The gas discs in both models are represented with $1\times10^6$ gas cells, and the stellar disc in the two-component model are represented with $5\times10^6$ $N$-body particles. The simulation code operates mesh regulations such as motions of gas cells, refinement and derefinement so that each gas cell keeps its initial mass within a factor of $2$. A gravitational softening length of a stellar particle is set to $\epsilon_{\rm s}=50~{\rm pc}$, and that of a gas cell varies as 2.5 times its approximated cell radius with the lower limit of $\epsilon_{\rm g,min}=50~{\rm pc}$; therefore gas contraction is limited to the scale $\sim\epsilon_{\rm g,min}$.

The other settings of our simulations are the same between the single- and two-component models. The whole halo regions of cubic volume of $200~{\rm kpc}$ on a side are filled with diffuse gas the density of which is $n_{\rm H}=10^{-6}~{\rm cm^{-3}}$; however the halo gas hardly affects our simulation results. Our simulations with the isothermal equation of state keeps the gas temperature at $10^4~{\rm K}$ independent of density. Haloes and bulges are represented with rigid potentials of Navarro-Frenk-White and Hernquist models, respectively \citep{nfw:97,h:90}. The masses of the halo and the bulge are $M_{\rm h}=1.1\times10^{12}~{\rm M_\odot}$\footnote{The value of $M_{\rm h}$ is defined to be the mass enclosed within the galactocentric radius $r_{\rm 200}=206~{\rm kpc}$ at which $3M_{\rm h}/(4\pi r_{\rm 200}^3)=2.9\times10^{4}{\rm M_{\odot}~kpc^{-3}}$ and is 200 times as large as the present-day cosmic background density.} and $M_{\rm b}=4.3\times10^{9}~{\rm M_\odot}$, and their scale radii are $r_{\rm h}=20.6~{\rm kpc}$ and $r_{\rm b}=0.3~{\rm kpc}$, respectively. Although we do not change the parameters of the halo and the bulge in this paper, our Paper I has demonstrated that our analysis is fairly robust and can characterise SAI independent of background potentials in our non-magnetic hydrodynamics/$N$-body simulations. 

Strength of magnetic fields is the most important parameter of this study. Using the initial conditions described above, we run the simulations with a parameter of $\beta_{\rm ini}\equiv c_{\rm snd}^2/v_{\rm A}^2=\infty$, $100$, $20$ and $5$.\footnote{Note that the definition of $\beta_{\rm ini}$ is not identical to that of $\beta$ in equation (\ref{ND_DR1}); the former does not take into account turbulent pressure of gas, whereas the latter does.} The value of $\beta_{\rm ini}$ is spatially uniform; since we assume the isothermal equation of state, all gas cells share the same $v_{\rm A}$ in the initial states. The magnetic fields are oriented azimuthally in the initial conditions and evolve self-consistently after starting the runs. Previous studies have performed MHD simulations with various initial conditions and methodology such as initially random orientation of magnetic fields in an isolated disc \citep{kk:18}, vertical orientation in a spherical collapse model for disc formation \citep{ps:13} and a uniform seed field in cosmological simulations \citep{pms:14,pgg:17,pgp:18}. However, independent of their initial conditions and types of simulations, these studies demonstrated that magnetic fields become nearly toroidal and oriented along spiral arms during galaxy formation and/or emergence of spiral arms.

\subsection{Data analysis}
\label{ana}
Following our Paper I, we employ the same analysis method based on polar maps. First, we apply two-dimensional Gaussian kernels whose full width at half maximum is $0.5~{\rm kpc}$ to gas and stars in snapshots. Then, we vertically integrate physical quantities weighted by mass and compute the quantities needed to solve equation (\ref{quadra} or \ref{sixthorder}): $\Sigma_{\rm g}$, $\Sigma_{\rm s}$, $\sigma_{\rm g}$, $\sigma_{\rm s}$, $\Omega$ and $v_{\rm A}$. Angular velocity is computed from local mean velocity as $\Omega=\overline{v_{\rm \phi}}/R$. In the two-component models, we compute $\Omega$ as the mass-weighted mean between gas and stellar components. The value of $v_{\rm A}$ in our analysis is computed using azimuthal component $B_\phi$ and local spatial density of gas. Hereafter we denote this `azimuthal Alfv\'en velocity' as $v_{{\rm A},\phi}$. We make polar plots of these quantities as functions of $(R,\phi)$ for each snapshot.

Our analysis also requires to detect spiral arms and measure their half widths $W_{\rm g}$ and $W_{\rm s}$. To this end, we use the same method proposed in our Paper I. In what follows, we describe the method briefly. In the polar plot of $\Sigma(R,\phi)$, we perform one-dimensional Gaussian fitting along the radial direction at a given $\phi$. The Gaussian fitting is iteratively adopted in the range from $R-W$ to $R+W$,\footnote{As we describe in Section \ref{basiceq}, $W=1.55w$ where $w$ is a deviation of a Gaussian function fitted.} while changing $W$. By computing a goodness-of-fit $\chi^2$ for each $W$, we search for the half width $W$ that gives the minimum value of $\chi^2$. Thus, we obtain the best-fit $W$ and its minimum $\chi^2$ at each coordinate point $(R,\phi)$. We perform this fitting procedure for each of gas and stellar component. The goodness-of-fit $\chi^2(R,\phi)$ is expected to become significantly lower than unity if there is a crest of a spiral arm at $R$ and if the radial distribution of $\Sigma$ is close to Gaussian. In this paper, we define spiral arms to be regions where $\log\chi_{\rm g}^2<-0.25$ for the single-component models, and $\log(\chi_{\rm g}^2+\chi_{\rm s}^2)<-0.1$ for the two-component models. Although this threshold of $\chi^2$ is arbitrary, our instability analysis to compute $s_{\rm grow}$ is independent of the threshold. Note that our method to measure $W$ assumes a pitch angle $\theta=0$, therefore the true width is overestimated by a factor of $1/\cos\theta$.

\section{Results}
\label{result}
\subsection{The single-component runs}
\label{GasDisc}
\begin{figure}
  \includegraphics[bb=0 0 760 807, width=\hsize]{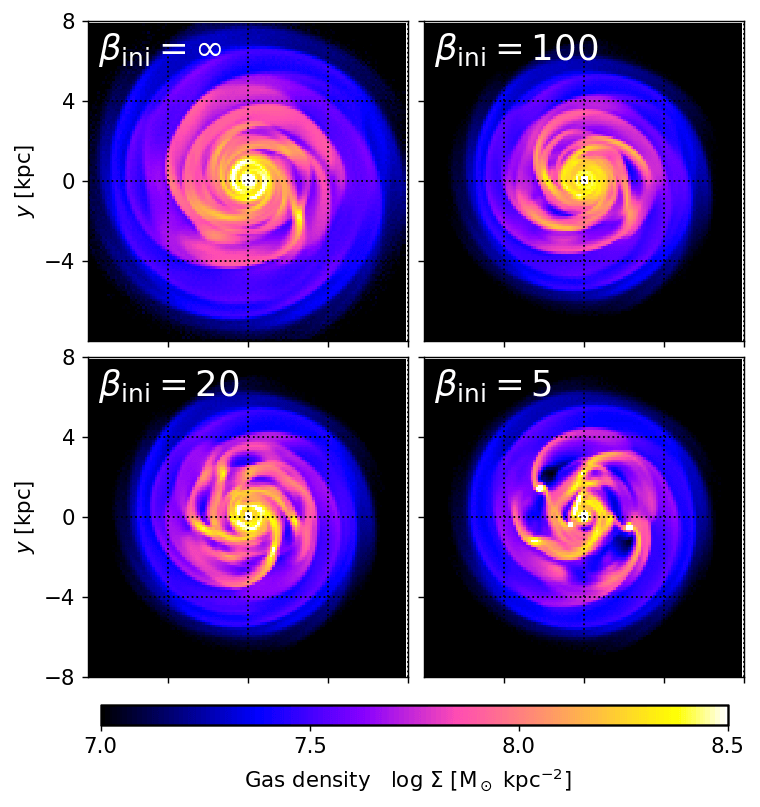}
  \caption{Gas surface densities at $t=400~{\rm Myr}$ in our single-component simulations with various $\beta_{\rm ini}$. The run with $\beta_{\rm ini}=\infty$ corresponds to the non-magnetic hydrodynamics simulation (top left). All runs start with the same initial conditions and settings except $\beta_{\rm ini}$.}
  \label{faces}
\end{figure}
In Fig. \ref{faces}, we show gas surface density distributions in our single-component runs with $\beta_{\rm ini}=\infty$, $100$, $20$ and $5$ at $t=400~{\rm Myr}$. In the absence of magnetic field ($\beta_{\rm ini}=\infty$, top left), spiral arms in the simulation are stable and do not fragment. As seen in the figure, the gas density distributions become clumpier for stronger magnetic fields with lower $\beta_{\rm ini}$. In the case of $\beta_{\rm ini}=100$ (top right), spiral arms are still stable at $t=400~{\rm Myr}$; however the density contrast between the arms and the inter-arm regions appears somewhat higher than in the non-magnetic run. Since any spiral arms are supposed to be weakly unstable in our linear analysis even in infinitesimal magnetic fields, perturbations can grow slowly (see Section \ref{ana_single}) although the arms in this run do not fragment in early stages of the simulation. Moreover, magnetic fields in spiral arms can be amplified because of accretion of gas coupled with magnetic fields onto the arms, transport of magnetic fields \citep{kk:18} and/or dynamo mechanisms by differential rotation of a galactic disc \citep{sss:06} and small-scale turbulence \citep{ssk:13}. The arms finally fragment at $t=800$--$900~{\rm Myr}$ even in the run with the weak magnetic fields ($\beta_{\rm ini}=100$).

In the case of $\beta_{\rm ini}=20$ (bottom left in Fig. \ref{faces}), gas densities vary significantly along spiral arms, and the arms have knotty structures. Although the arms in this run may be expected to be unstable and to fragment in the snapshot, they actually do not fragment or form massive clumps soon after this snapshot. This run finally shows fragmentation of a spiral arm at $t=600$--$700~{\rm Myr}$.

 In the case of $\beta_{\rm ini}=5$ (bottom right in Fig. \ref{faces}), our simulation clearly shows that spiral arms form massive clumps via fragmentation at $t=400~{\rm Myr}$. We regard the arms in this run to be strongly unstable. 

\begin{figure}
  \includegraphics[bb=0 0 971 1051, width=\hsize]{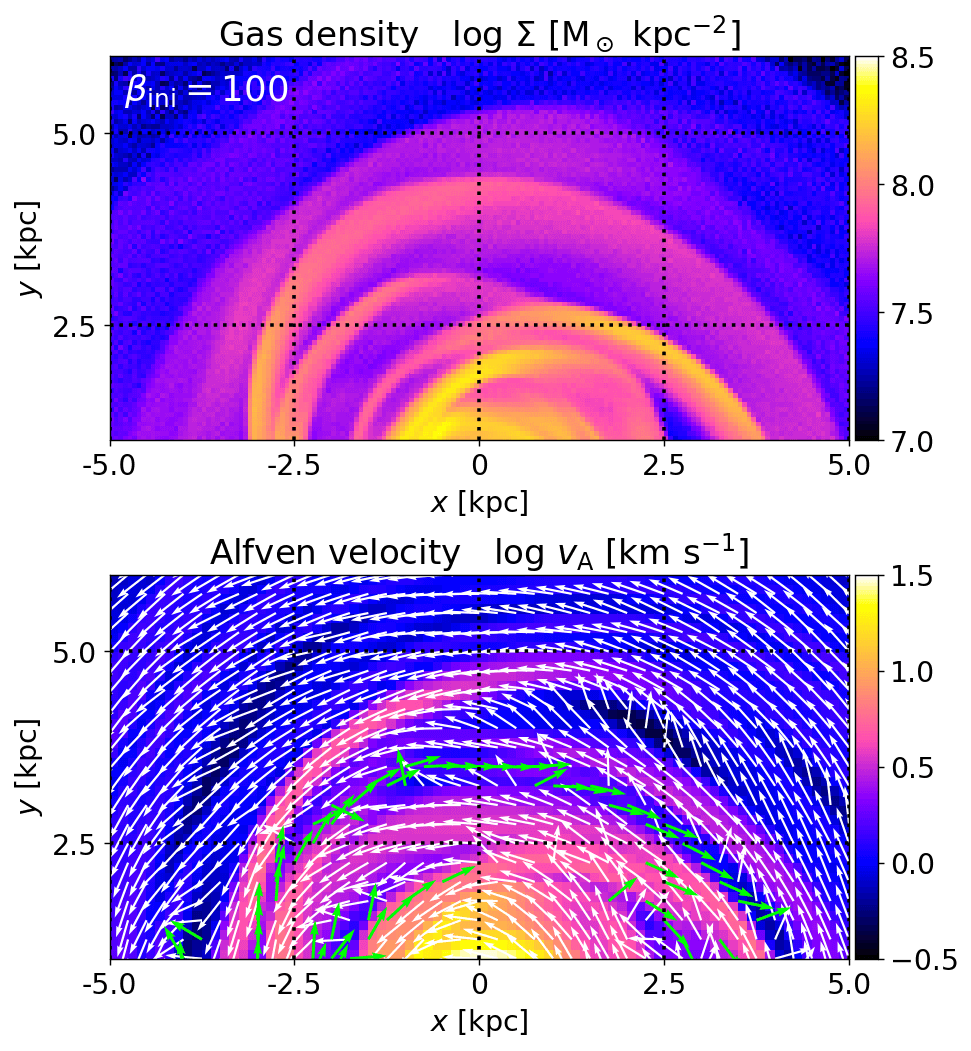}
  \caption{Distribution of gas densities (top) and Alfv\'en velocities (bottom) in the single-component run with $\beta_{\rm ini}=100$ at $t=400~{\rm Myr}$. The top panel is the same as the top right panel of Fig. \ref{faces}. In the bottom panel, arrows indicate orientation of magnetic fields, where the white arrows correspond to co-rotating magnetic fields ($B_\phi>0$), and the green ones indicate field reversals ($B_\phi<0$). Here the Alfv\'en velocities are computed from all components of \mathbfit{B}, as $v_{\rm A}=|\mathbfit{B}|/\sqrt{4\pi\rho}$.}
  \label{B1}
\end{figure}
Fig. \ref{B1} shows gas densities (top) and strength of magnetic fields represented with Alfv\'en velocities (bottom) in the run with $\beta_{\rm ini}=100$ at $t=400~{\rm Myr}$. Clear correlation between the two can be seen. High density regions such as spiral arms generally have strong magnetic fields. In the bottom panel, the magnetic fields are basically oriented along the spiral arms, whereas they are oriented counter-rotating (the green arrows) with $B_\phi<0$ in some inter-arm regions. Such field orientations between arms and inter-arm regions have also been observed in the Milky Way \citep[][and references therein]{h:17}. These trends, such as correlation of field strength with gas density, nearly toroidal magnetic fields following spiral arms and field reversals in inter-arm regions, are generally seen in our MHD simulations. However, we also see that spiral arms can occasionally have significant radial component of magnetic fields (see Section \ref{radialB}).

\subsubsection{The case of the strong magnetic fields}
\label{unstable}
\begin{figure*}
	\begin{minipage}{\hsize}
		\includegraphics[bb=0 0 1696 1006,width=\hsize]{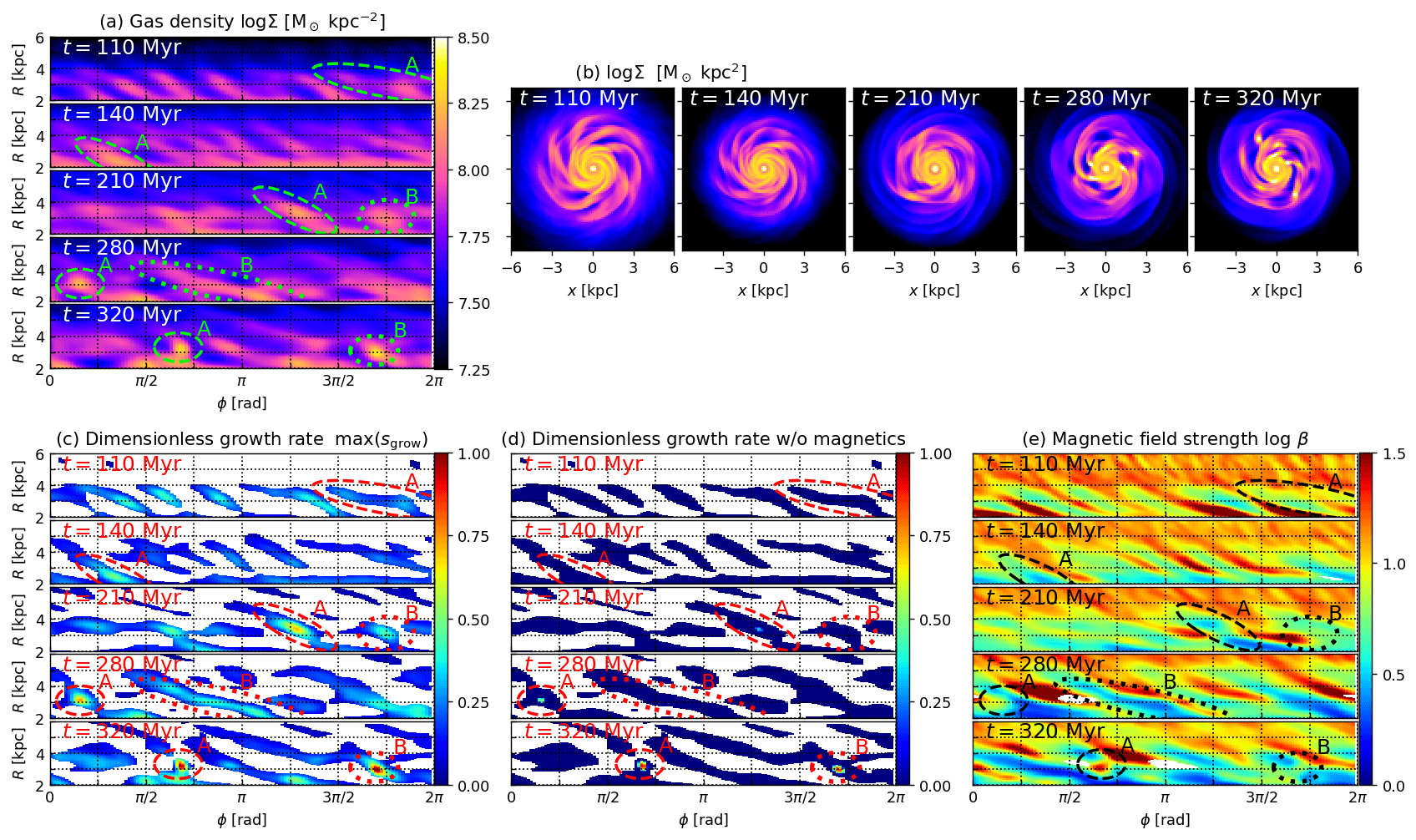}
		\caption{Polar-map analysis of the single-component run with $\beta_{\rm ini}=5$ at $t=110$, $140$, $210$, $280$ and $320~{\rm Myr}$. \textit{Panels a and b}: gas surface density distributions in the polar and the face-on Cartesian coordinates. \textit{Panel c}: the maximum values of dimensionless growth rates $\max[s_{\rm grow}(x)]$ computed from our single-component analysis, where regions with high $\max(s_{\rm grow})$ are expected to be unstable in the linear analysis. \textit{Panel d}: the same as Panel c but ignoring the effect of magnetic fields. \textit{Panel e}: strength of toroidal magnetic fields, where $\beta\equiv (c_{\rm snd}^2+\sigma_{\phi}^2)/v_{{\rm A},\phi}^2$. In Panels c and d, the inter-arm regions are uncoloured, where our arm-detection method indicates $\log\chi^2>-0.25$ in the Gaussian fitting for the gas density distribution. In Panel e, regions of field reversals with $B_\phi<0$ are uncoloured. The dashed ellipses labelled as `A' and `B' in each panel trace the unstable spiral arms we focus on. The regions A and B fragment into gas clumps at $t=280$ and $320~{\rm Myr}$, and both regions indicate high values of $\max[s_{\rm grow}(x)]\gsim0.3$ in our linear analysis before the fragmentation (Panel c). On the other hand, the analysis ignoring the magnetic effect cannot capture the fragmentation (Panel d).}
		\label{Unstable}
	\end{minipage}
\end{figure*}
In the run with the strong magnetic fields with $\beta_{\rm ini}=5$, the spiral arms fragment and form clumps within a few orbital time-scales after the simulation is started. Here, we focus on the unstable states of the fragmenting arms, which are expected to indicate high values of $\max(s_{\rm grow})$ in our linear analysis. Fig. \ref{Unstable} shows our polar-map analysis for the snapshots at $t=110$, $140$, $210$, $280$ and $320~{\rm Myr}$. In these snapshots, the dashed ellipses labelled as `A' and `B' in the figure indicate the two spiral arms that are fragmenting. Panels a and b show gas surface densities in the polar and Cartesian coordinates. Regions A and B fragment and form gas clumps at $t=280$ and $320~{\rm Myr}$, respectively. 

Panel c of Fig. \ref{Unstable} shows distributions of $\max(s_{\rm grow})$ computed from our single-component linear analysis described in Section \ref{ana_single}. At $t=140$ and $210~{\rm Myr}$, although the spiral arms in the regions A and B have not fragmented yet, they indicate high values of $\max(s_{\rm grow})\simeq0.3$. Indeed, in the following snapshots at $t=280$ and $320~{\rm Myr}$, the arms fragment and form the clumps in the regions A and B. At $t=110~{\rm Myr}$, although the spiral arms other than Region A also indicate $\max(s_{\rm grow})\simeq0.3$, the density peaks in these arms quickly migrate into the galactic centre along the arms. After the migration of the density peaks, the remaining arms have low $\Sigma$ and indicate $\max(s_{\rm grow})$ smaller than before the migration. They do not fragment. Panel d shows $\max(s_{\rm grow})$ ignoring the effect of magnetic fields, computed from the same linear analysis but assuming $\beta=\infty$. The non-magnetic analysis indicates $s_{\rm grow}=0$ for the fragmenting spiral arms and $s_{\rm grow}>0$ only inside the clumps after their collapse. Thus, the linear analysis ignoring magnetic fields cannot characterise the fragmentation correctly and underestimates the instability since it does not taking into account the magnetic destabilisation effect. Panel e shows strength of toroidal magnetic fields $\beta$ computed using $v_{\rm A,\phi}$ in the simulation.\footnote{Here, we take into account thermal and turbulent pressure in computing $\beta\equiv(c_{\rm snd}^2+\sigma_{\phi}^2)/v_{{\rm A},\phi}^2$.} In this run, the fragmenting spiral arms have approximately $\log\beta\simeq0.5$. In Appendix \ref{wavelength}, we show physical length scales of the perturbations with respect to arm widths in this run, and discuss applicability of our analysis.

\subsubsection{The  case of the weak magnetic fields}
\label{singlestable}
\begin{figure*}
	\begin{minipage}{\hsize}
		\includegraphics[bb= 0 0 1696 657,width=\hsize]{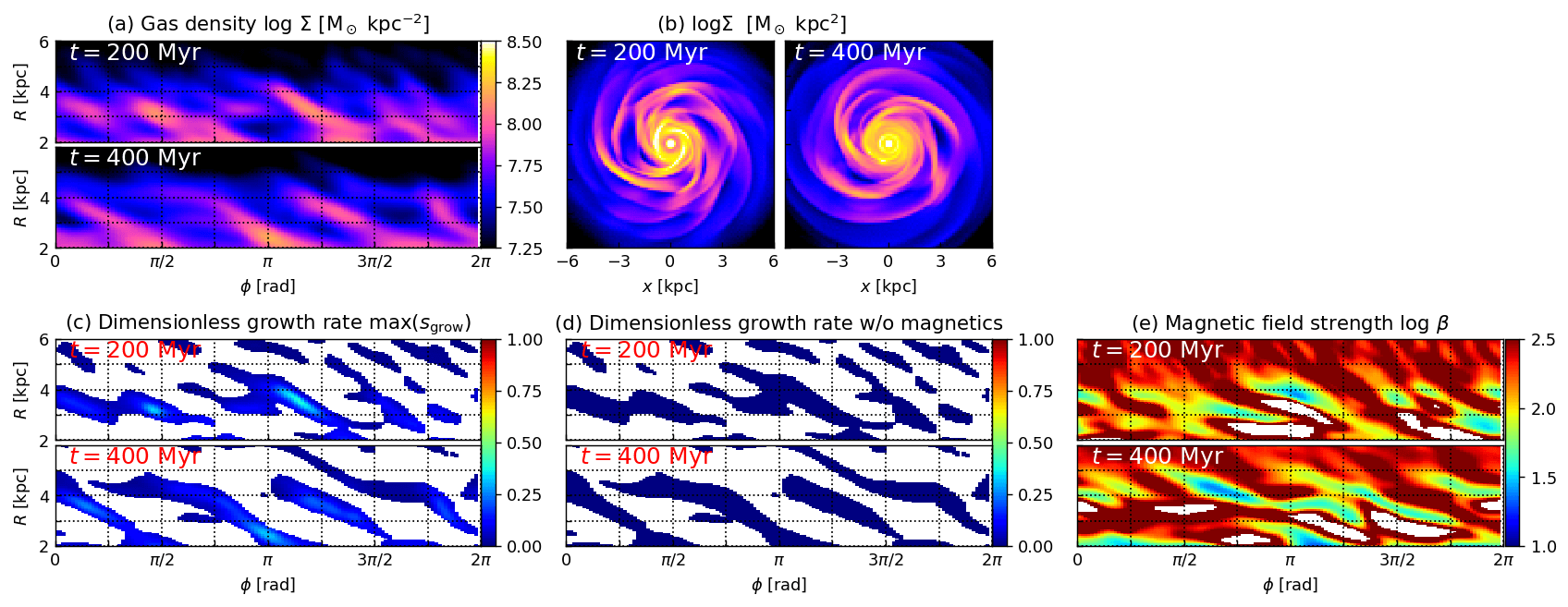}
		\caption{Same as Fig. \ref{Unstable} but for the single-component run with $\beta_{\rm ini}=100$ at $t=200$ and $400~{\rm Myr}$. In this run, no spiral arms fragment, and  $\max(s_{\rm grow})$ indicates low values $\lsim0.3$ in Panel c.}
		\label{Stable}
	\end{minipage}
\end{figure*}
The spiral arms in our single-component run with the weak magnetic fields of $\beta_{\rm ini}=100$ do not fragment until $t\simeq800$--$900~{\rm Myr}$ (the top right panel of Fig. \ref{faces}). Here we focus on the arms that are only weakly unstable before the fragmentation. We expect that our instability analysis suggests low values of $s_{\rm grow}$ inside the spiral arms in early stages of the run. Fig. \ref{Stable} shows the result of our linear analysis for this stable run at $t=200$ and $400~{\rm Myr}$. Panels a and b show gas surface densities, where no fragmentation is seen. Panel c indicates distributions of $\max(s_{\rm grow})$, where the growth rates are $\lsim0.3$ and significantly lower than in the case of the strong magnetic fields (Panel c in Fig. \ref{Unstable}) although a spiral arm indicates somewhat high values of $\max(s_{\rm grow})\simeq0.3$ at $t=200~{\rm Myr}$ (see Section \ref{radialB}). The arms with these low values of $\max(s_{\rm grow})\lsim0.3$ seen in this run are considered to be practically stable over a disc-rotation time-scale. Low $\max(s_{\rm grow})$ corresponds to weakly unstable states in linear regime where perturbations can grow slowly. In such cases, effects that are not taken into account in our analysis can play important roles in the dynamics of spiral arms and may prevent the perturbations from growing; for example, since pitch-angles of the arms are not exactly $\theta=0$, differential rotation of the disc may stretch the arms and decrease the densities within them.

Panel d indicates $\max(s_{\rm grow})$ but ignoring the magnetic field effects. The analysis predicts no instability, i.e. $s_{\rm grow}=0$ for all perturbations. Panel e in Fig. \ref{Stable} indicates the strength of magnetic fields $\beta$. In this run with $\beta_{\rm ini}=100$, the magnetic fields are significantly weaker than in the run with $\beta_{\rm ini}=5$ shown in Fig. \ref{Unstable} (note the different colour scales between Panels e in Figs. \ref{Unstable} and \ref{Stable}). The spiral arms in this run have $\log\beta\simeq1.0$--$1.5$. Because the other properties such as surface densities and widths of the spiral arms are not largely different between the cases with $\beta=5$ and $100$, the early fragmentation seen in the former run can be attributed to the strong magnetic fields, i.e. the magnetic destabilisation.

\subsubsection{The  case of the moderate magnetic fields}
\label{nonlinear}
\begin{figure*}
	\begin{minipage}{\hsize}
		\includegraphics[bb=0 0 1696 853,width=\hsize]{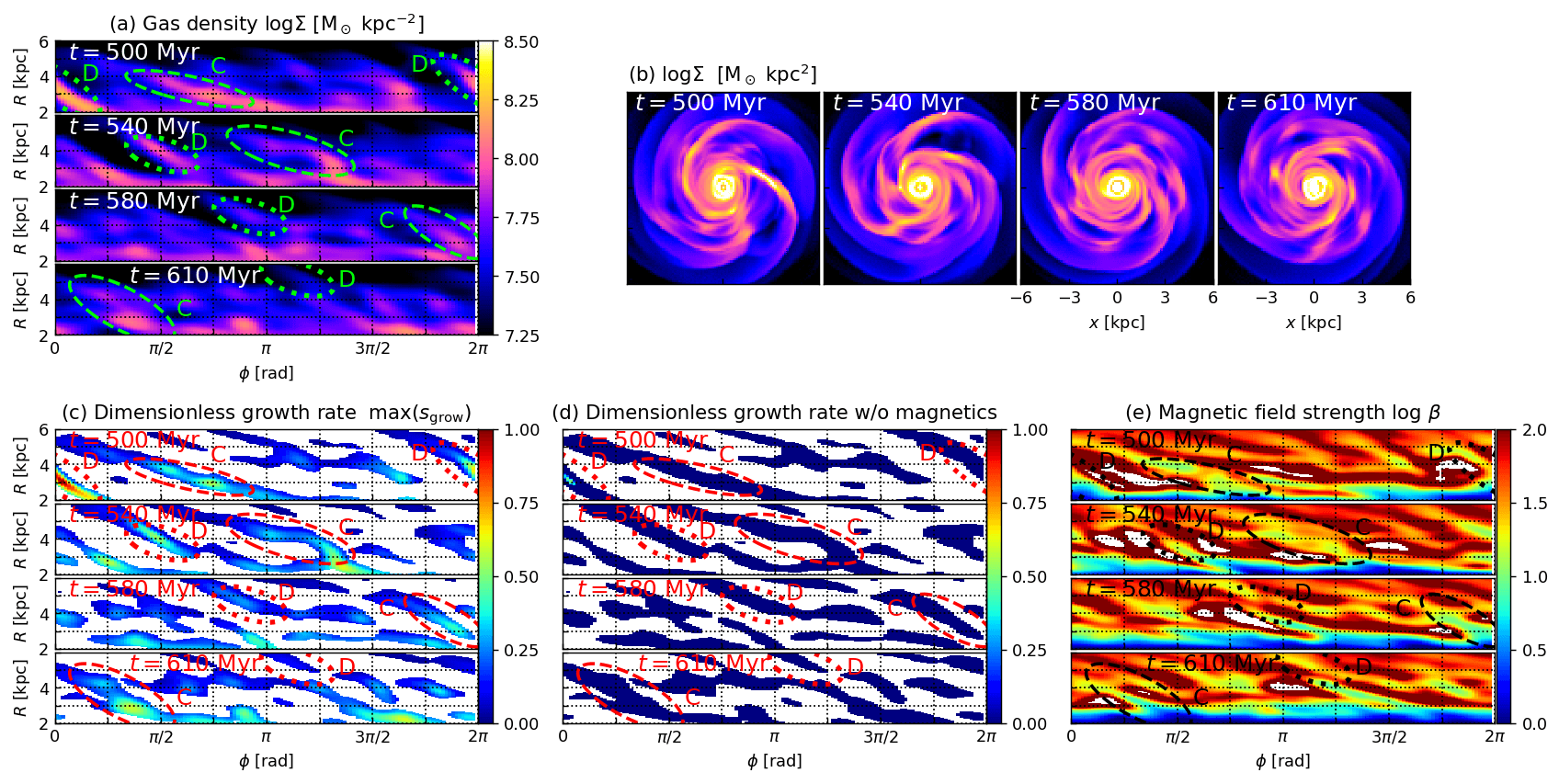}
		\caption{Same as Fig. \ref{Unstable} but for the single-component run with $\beta_{\rm ini}=20$ at $t=500$, $540$, $580$ and $610~{\rm Myr}$. In this run, the spiral arms do not clearly fragment or form giant clumps. Although the regions labelled as `C' and `D' indicate high values of $\max(s_{\rm grow})\gsim0.3$ in the linear analysis, these regions actually appear to be stable in the simulation.}
		\label{Nonlinear}
	\end{minipage}
\end{figure*}
Fig. \ref{Nonlinear} shows results of our linear analysis applied to the run with the moderate strength of magnetic fields of $\beta_{\rm ini}=20$. As seen in Panels a and b, the spiral arms in this run do not clearly fragment or form clumps even in later snapshots at $t\simeq500$--$600~{\rm Myr}$ although the arms may appear clumpy and to be fragmenting at $t=400~{\rm Myr}$. In Fig. \ref{Nonlinear}, we focus on two spiral arms that are labelled as `C' and `D'. In Panel c, these labelled regions indicate high values of $\max(s_{\rm grow})\simeq0.5$ and $0.8$ at $t=500~{\rm Myr}$. However, the arms do not form clumps and appear to be stable over an orbital time-scale. The Region C keeps the high values around $\max(s_{\rm grow})\simeq0.3$--$0.5$, and the region D decreases $\max(s_{\rm grow})$ in the following snapshots. The arm in the region D is finally dissolved at $t=610~{\rm Myr}$. Thus, in this run, although the spiral arms are supposed to be unstable in our linear analysis, they are actually stable in the simulation. We argue this inconsistency in Section \ref{radialB}.

Panel d in Fig. \ref{Nonlinear} shows dimensionless growth rates ignoring magnetic fields. We find $\max(s_{\rm grow})=0$ except the region D at $t=500~{\rm Myr}$; therefore the high values of $\max(s_{\rm grow})$ seen in Panel c are due to the magnetic fields. In Panel e, the spiral arms have approximately $\log\beta\simeq0.5$--$1.0$ showing that the magnetic fields are stronger than in the stable case shown in Fig. \ref{Stable}.

\subsubsection{Effects of radial magnetic fields}
\label{radialB}
\begin{figure*}
  \includegraphics[bb=0 0 2671 790, width=\hsize]{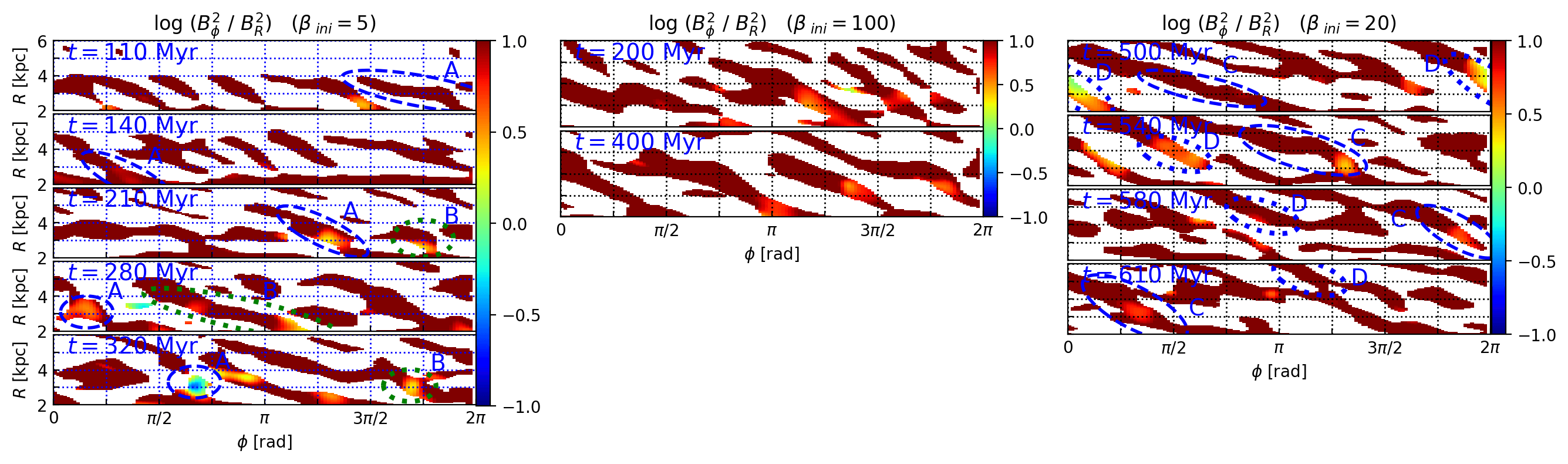}
  \caption{Ratios of $\log(B_\phi^2/B_R^2)=\log(v_{{\rm A},\phi}^2/v_{{\rm A},R}^2)$ in the spiral-arm regions in the single-component runs with $\beta_{\rm ini}=5$ (left), $100$ (middle) and $20$ (right).  The snapshots shown here are the same as those in Figs. \ref{Unstable}, \ref{Stable} and \ref{Nonlinear}. In Arm D, the magnetic fields have significant radial components: $B_\phi\simeq B_R$.}
  \label{vAratio}
\end{figure*}
Because our analysis relies on various assumptions, its applicability may not hold in some cases. Here we discuss limitations of our analysis. In Arm D (Fig. \ref{Nonlinear}) in the single-component run with $\beta_{\rm ini}=20$, the stable state and the following dissolution in the simulation is remarkably discrepant with our analysis indicating the high $\max(s_{\rm grow})\simeq0.8$ at $t=500~{\rm Myr}$ (Panel c). Here we discuss the reason why our analysis fails for Arm D. Fig. \ref{vAratio} shows ratios of $B_\phi^2/B_R^2=v_{{\rm A},\phi}^2/v_{{\rm A},R}^2$ in the same snapshots of the runs with $\beta_{\rm ini}=5$ (left), $100$ (middle) and $20$ (right). At $t=500~{\rm Myr}$ in the right panel, magnetic fields in Arm D have a strong radial component: $B_R\simeq B_\phi$. However, our analysis cannot take into account effects of the radial fields. Radial component of the fields can exert magnetic force in the azimuthal direction and is expected to suppress perturbations along an arm. Thus, we infer that the stable state, albeit with high $\max(s_{\rm grow})$, could be attributed to the strong radial magnetic fields in Arm D. Region C in the same run ($\beta_{\rm ini}=20$) also has radial magnetic fields in $t\geq540~{\rm Myr}$: $B_\phi^2/B_R^2\simeq3$--$5$. Although it is not clear how influential the radial fields are in Region C, the stable state with $\max(s_{\rm grow})\simeq0.3$--$0.5$ in Region C may also be due to the radial magnetic fields.

In the run with the weak magnetic fields ($\beta_{\rm ini}=100$, Fig. \ref{Stable}), we see the spiral arm that does not fragment but has $\max(s_{\rm grow})\simeq0.3$ at $t=200~{\rm Myr}$. The ratios in the arm are shown in the middle panel of Fig. \ref{vAratio} and are $B_\phi^2/B_R^2\gsim3$. As we discussed above, we infer that this arm might also be affected by the radial fields and prevented from fragmenting even with $\max(s_{\rm grow})\simeq0.3$. In the run with the strong magnetic fields ($\beta_{\rm ini}=5$, Fig. \ref{Unstable}), magnetic fields are almost toroidal with $B_\phi^2/B_R^2\gsim10$, except inside the clumps that have already collapsed. Hence, we expect that our analysis can characterise the fragmentation in this run.

\subsubsection{An instability criterion}
\label{Criterion}
From the results of our simulations with the various $\beta_{\rm ini}$ and the discussion above, we consider that spiral arms with $\max(s_{\rm grow})\gsim0.3$ would fragment and collapse into clumps within an orbital time-scale. However, it should be noted that our analysis becomes inapplicable if there are significant radial magnetic fields, as we showed above. It should be noted that behaviours of magnetic fields in disc simulations are quite complicated as shown in Fig. \ref{B1}, and radial fields are not necessarily negligible in spiral arms even if the fields are toroidal in the initial conditions. Therefore, when our magnetic SAI analysis is adopted to observed spiral galaxies, accurate determinations of magnetic fields are needed.

Our linear perturbation analysis described in Section \ref{basiceq} also posits on the assumptions such as equilibrium states and rigid rotations in spiral arms wound tightly. Deviation from these assumptions could cause various effects that are not considered in our analysis. In Section \ref{othereffects}, we discuss such non-linear effects and other physics that our linear analysis lacks. These effects can possibly stabilise and prevent arms from fragmenting when simulations deviate from the linear analysis and break the assumptions in our linear analysis.


\subsection{The two-component runs}
\label{2compDisc}
\begin{figure}
  \includegraphics[bb=0 0 425 453, width=\hsize]{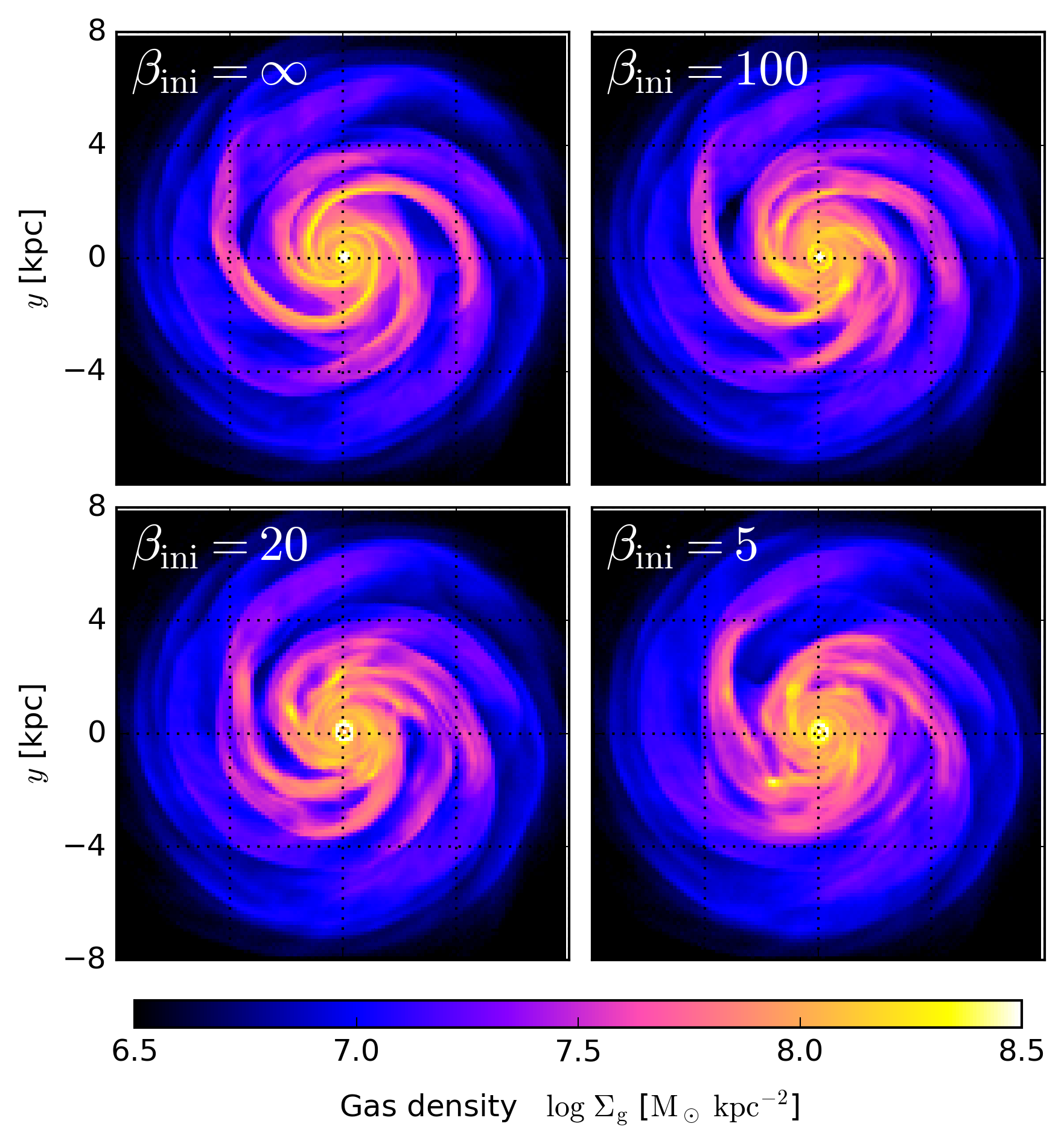}
  \includegraphics[bb=0 0 425 453, width=\hsize]{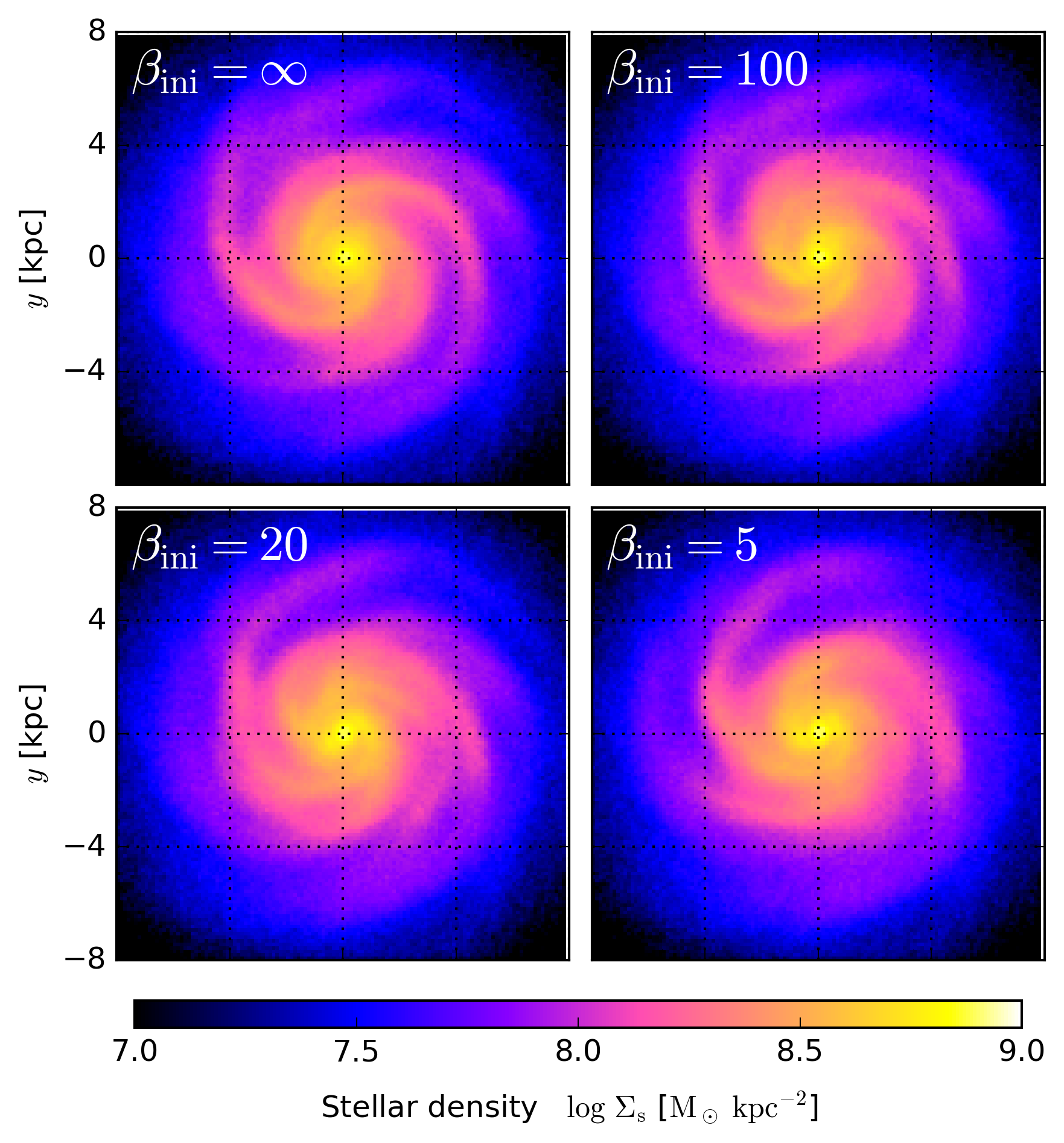}
  \caption{Distributions of surface densities at $t=300~{\rm Myr}$ in our simulations with various $\beta_{\rm ini}$. Top and bottom sets of panels show gas and stellar densities. All runs start with the same initial conditions and settings except $\beta_{\rm ini}$.}
  \label{faces2}
\end{figure}
In Fig. \ref{faces2}, we show surface density distributions of gas (top set of panels) and stellar  (bottom set of panels) components in our two-component simulations at $t=300~{\rm Myr}$, in which the initial magnetic field strength is $\beta_{\rm ini}=\infty$, $100$, $20$ and $5$. Spiral-arm structures can be seen in both gas and stellar distributions, and we adopt our two-component linear analysis described in Section \ref{2compana} to these simulations.

In the non-magnetic run with $\beta_{\rm ini}=\infty$, the spiral arms are stable at least until $t=1~{\rm Gyr}$. Similarly to the single-component runs (Fig. \ref{faces}), the galaxies become clumpy and their arms tend to have knotty structures as $\beta_{\rm ini}$ decreases. Even in the case of the strong magnetic field with $\beta_{\rm ini}=5$, however, the clumpy structures in gas do not appear to be massive enough to capture stellar particles; stellar clumps are not seen in the surface density maps of the stellar components. In Fig. \ref{faces2}, such low-mass gas clumps could not completely tear the spiral arms. Thus, the magnetic destabilisation may be limited in the two-component runs even if $\beta_{\rm ini}=5$. This would be because magnetic fields only affect the gas component accounting for the small fraction ($f_{\rm g}=0.175$) of the total disc mass, and the stars dominate dynamics within the discs.

\begin{figure*}
	\begin{minipage}{\hsize}
		\includegraphics[bb=0 0 2621 1330,width=\hsize]{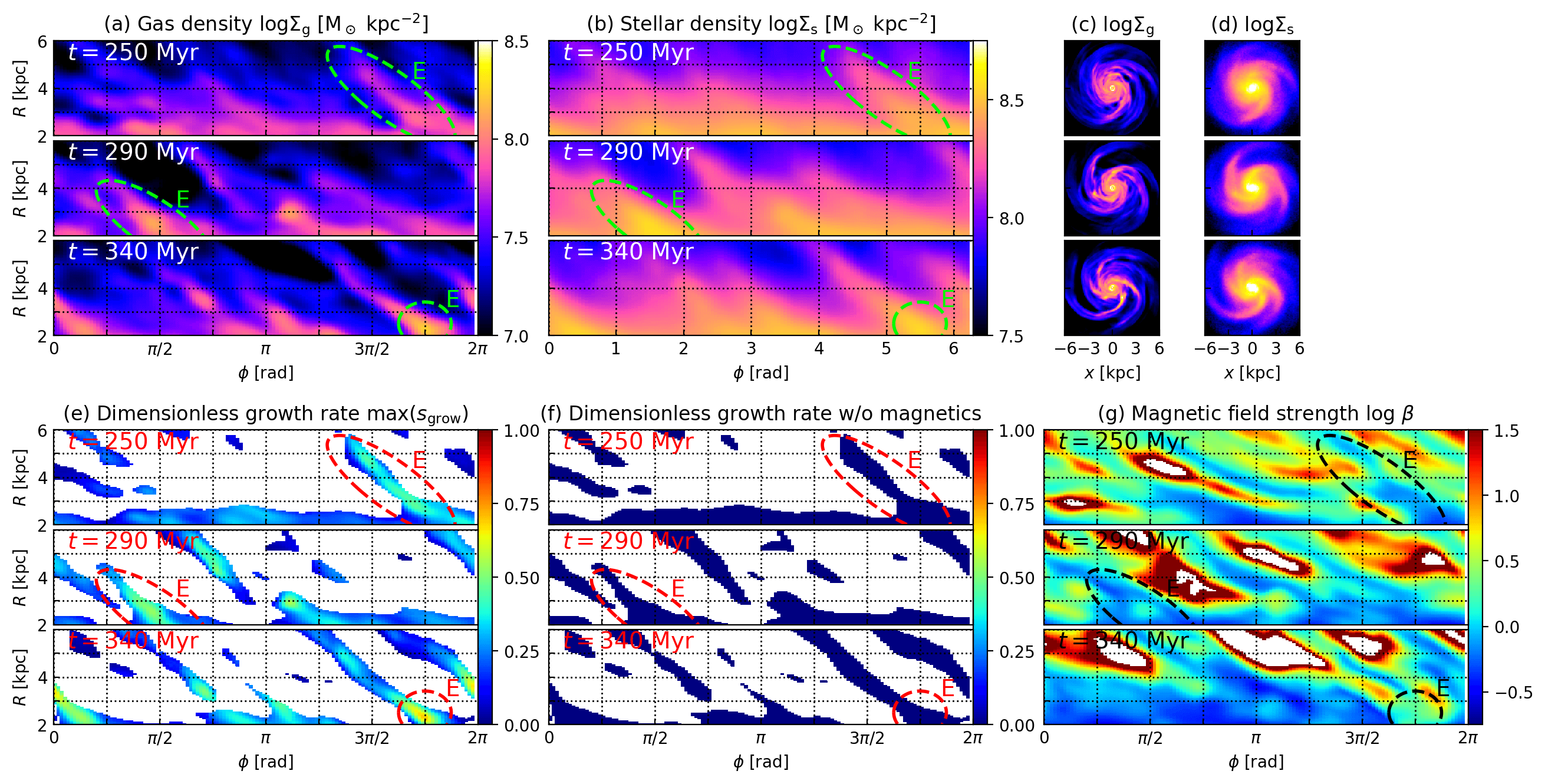}
		\caption{Polar-map analysis of the two-component run with $\beta_{\rm ini}=5$ at $t=250$, $290$ and $340~{\rm Myr}$. \textit{Panels from a to d}: surface density distributions of gas and stars. \textit{Panel e}: the maximum values of dimensionless growth rates $\max[s_{\rm grow}(k)]$ computed from our two-component analysis. The inter-arm regions are uncoloured, where our arm-detection method indicates $\log(\chi_{\rm g}^2+\chi_{\rm s}^2)>-0.1$ in the Gaussian fittings. \textit{Panel f}: same as Panel e, but ignoring the magnetic fields. \textit{Panel g}: strength of toroidal magnetic fields, where the regions with field reversals $B_\phi<0$ are uncoloured. In Panels a and b, the region labelled as `E' appears to be unstable and form a gas clump at $t=340~{\rm Myr}$ in the simulation although the arm is not completely torn off. In Panel e, our linear analysis also indicates high $\max[s_{\rm grow}(k)]>0.3$ in Region E before the collapse. If we ignore the magnetic fields in Panel f, on the other hand, our linear analysis indicates $\max[s_{\rm grow}(k)]=0$ and cannot capture the instability.}
		\label{Unstable2}
	\end{minipage}
\end{figure*}
In the two-component run with $\beta_{\rm ini}=5$, a few small gas clumps can be seen at $t=300~{\rm Myr}$ in the bottom right panel of Fig. \ref{faces2}. Fig. \ref{Unstable2} shows our polar-map analysis for the snapshots at $t=250$, $290$ and $340~{\rm Myr}$, where a fragmenting arm is marked as `E' with dashed ellipses. Since Arm E is not completely torn off and keeps the spiral structure without destructing the arm even after the gas clump formation, we consider that this arm is weakly fragmenting. Eventually, the clump forming in Arm E migrates along the spiral arm to the galactic centre. Panel e of Fig. \ref{Unstable2} shows the dimensionless growth rates $\max(s_{\rm grow})\simeq0.4$--$0.5$ in the fragmenting arm, which indicate rapid growth of perturbations and are consistent with the result of the formation of the gas clump in the following snapshots. In Panel d, we find that, if we ignore magnetic fields in our linear analysis, the arm is predicted to be stable with $s_{\rm grow}=0$ for all perturbations in these snapshots. The non-magnetic analysis thus cannot characterise the fragmentation. In Panel g, we show distributions of magnetic field strength $\beta$, and the fragmenting arm in this run have quite low $\log\beta\simeq-0.3$--$0$. The low values $\log\beta\sim0$ suggest energy equipartition between (thermal and turbulent) pressure and toroidal magnetic field. Despite the strong magnetic fields, the arm is only weakly unstable in this two-component run. In disc galaxies with such subdominant gas components, the magnetic destabilisation appears to be limited even if energy equipartition is reached.

\begin{figure*}
	\begin{minipage}{\hsize}
		\includegraphics[bb= 0 0 2648 1024,width=\hsize]{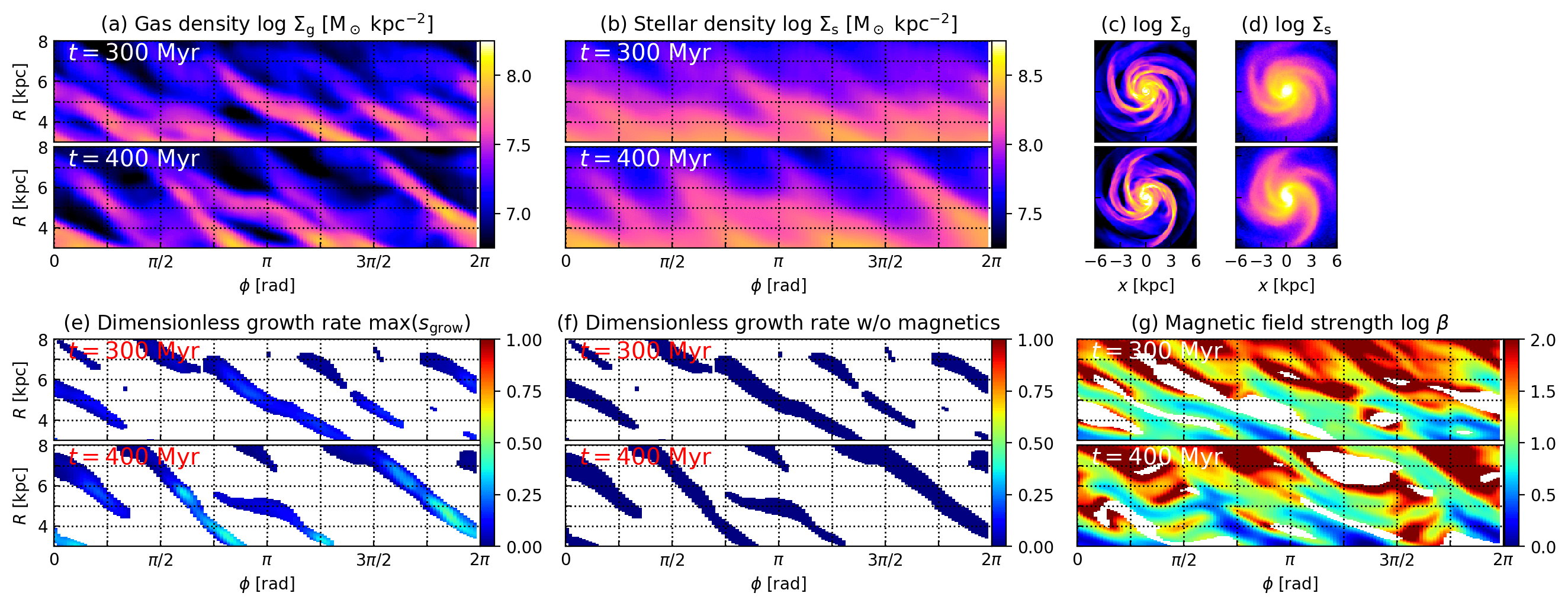}
		\caption{Same as Fig. \ref{Unstable2} but for the two-component run with $\beta_{\rm ini}=100$ at $t=300$ and $400~{\rm Myr}$. The spiral arms do not fragment or form clumps until $t\simeq500~{\rm Myr}$, and our linear analysis indicates low values of $\max(s_{\rm grow})\lsim0.3$ in Panel f.}
		\label{Stable2}
	\end{minipage}
\end{figure*}
The two-component run with $\beta=100$ (the top right panels in Fig, \ref{faces2}) does not form clumps until $t\sim500~{\rm Myr}$ although there are density fluctuations in spiral arms after $t\sim300~{\rm Myr}$. In Fig. \ref{Stable2}, we show our polar-map analysis for the snapshots at $t=300$ and $400~{\rm Myr}$. Panel e shows that low values of $\max(s_{\rm grow})\lsim0.2$ in all spiral arms at $t=300~{\rm Myr}$, which can be considered to be practically stable states for the arms. At $t=400~{\rm Myr}$, however, the growth rates increase to $\max(s_{\rm grow})\simeq0.3$ for the arms although they do not fragment at least during a few orbital time-scales after this snapshot. We consider, therefore, that the spiral arms in this run might be stabilised by non-linear effects and/or physical mechanisms missing in our analysis such as radial magnetic fields (see Section \ref{radialB}). Panel f shows dimensionless growth rates but ignoring the magnetic fields, indicates stable states with $\max(s_{\rm grow})=0$. Panel g shows strength of the magnetic fields. The spiral arms have $\log\beta\simeq0.5$--$1.0$ approximately, and it appears that the values of $\beta$ decrease with time between $t=300$ and $400~{\rm Myr}$. The magnetic fields are thus amplified in the arms during the run.

In our two-component simulations, stable spiral arms with $\max(s_{\rm grow})\gsim0.3$ are often seen, such as the above case with $\beta=100$ at $t=400~{\rm Myr}$. We find that these arms with high $\max(s_{\rm grow})$ indicate significant radial fields. These stable arms are expected to be stabilised by the radial fields and beyond the applicable domain of our analysis as we discussed in Section \ref{radialB}. Again it should be noted that our SAI analysis including magnetic destabilisation has to be carefully applied with such complexity of magnetic fields.

\section{Discussion}
\label{discussion}
\subsection{Influence on clump mass}
\label{ClumpMassDiscuss}
Using our linear perturbation theory of SAI presented in Section \ref{basiceq}, we discuss a physical property of giant clumps observed in gas-rich star-forming galaxies. In our Paper I, we have demonstrated that our linear analysis can be applied to estimating masses of clumps forming via spiral-arm fragmentation in non-magnetised isolated galaxy simulations. In observations, the large masses of giant clumps, $M_{\rm cl}\lsim10^8~{\rm M_\odot}$, are most noticeable difference from normal star clusters and giant molecular clouds in spiral galaxies. Here, we discuss how toroidal magnetic fields can affect a typical mass of gas clumps forming via SAI. Because clumpy galaxies are generally observed to be highly gas-rich, we can assume that SAI is driven by gas in these galaxies (Paper I). In the following analysis, we therefore consider the single-component analytic model described in Section \ref{ana_single}. Hereafter, all variables such as $\Upsilon$, $\Sigma$, $\sigma$ and $W$ represent the physical properties of gas in a spiral arm.

Spiral-arm fragmentation is considered to occur in a marginally unstable state with a low value of $\max(s_{\rm grow})$. From the results of our single-component simulations, we estimate that the critical $\max(s_{\rm grow})$ for clump formation would be $\simeq0.3$ (Section \ref{Criterion}). Once the instability criterion is determined, using the analysis shown in Fig. \ref{DRall}, we can obtain $Q_{\rm sp}=2A\sigma\Omega W/(\pi G\Upsilon)$ of the critical states, at given $q\equiv\sigma/(2\Omega W)$ and $\beta$. In addition, by numerically solving the dispersion relation (\ref{ND_DR2}), we can also compute dimensionless wavenumber $x_{\rm os}\equiv k_{\rm os}W$ that gives $\max(s_{\rm grow})$ for the critical states. The onset of instability is expected to occur at the most unstable perturbation $k_{\rm os}$, and the perturbation grows and collapses first.

When the wavelength of the instability onset $\lambda_{\rm os}=2\pi/k_{\rm os}$ is significantly longer than a width of the arm (i.e. $x_{\rm os}\ll1$), the unstable perturbation can be expected to collapse along the spiral arm: one-dimensional collapse (large-scale SAI). In this case, the typical mass of a clump forming via SAI is estimated to be 
\label{clumpmass}
\begin{equation}
M_{\rm cl,1D}\sim\Upsilon\lambda_{\rm os}=8A\frac{\Omega^2W^3}{G}\frac{q}{Q_{\rm sp}x_{\rm os}},
\label{mcl1d}
\end{equation}
where again $\Upsilon=AW\Sigma$. Thus, $M_{\rm cl,1D}\propto q/(Q_{\rm sp}x_{\rm os})$ for given $\Omega$ and $W$, and we define a `clump-mass module' as $m_{\rm cl,1D}\equiv q/(Q_{\rm sp}x_{\rm os})$.

When $\lambda_{\rm os}$ is significantly shorter than the arm width (i.e. $x_{\rm os}\gg1$), the unstable perturbation is deeply embedded within the spiral arm. In this case, we consider that a round region with a radius of $\lambda_{\rm os}/2$ collapses: two-dimensional collapse (small-scale SAI).\footnote{This small-scale SAI reduces to Toomre instability against azimuthal perturbations within a spiral arm (Paper I).} In this case, we estimate the typical clump mass as
\begin{equation}
M_{\rm cl,2D}\sim\pi\Sigma\left(\frac{\lambda_{\rm os}}{2}\right)^2=4\pi^2\frac{\Omega^2W^3}{G}\frac{q}{Q_{\rm sp}x^2_{\rm os}},
\label{mcl2d}
\end{equation}
and a clump-mass module is defined as $m_{\rm cl,2D}\equiv q/(Q_{\rm sp}x_{\rm os}^2)$. \footnote{However, note that an unstable perturbation would collapse three-dimensionally if $\lambda_{\rm os}$ is shorter than a vertical height of the spiral arm. In this case, a resultant clump mass can become smaller than $M_{\rm cl,2D}$.} We assume these clump-mass moduli, $m_{\rm cl,1D}$ and $m_{\rm cl,2D}$, to switch at $x_{\rm os}=1$ although the boundary between the two regimes is ambiguous in reality.

\begin{figure}
  \includegraphics[bb=0 0 2845 2463, width=\hsize]{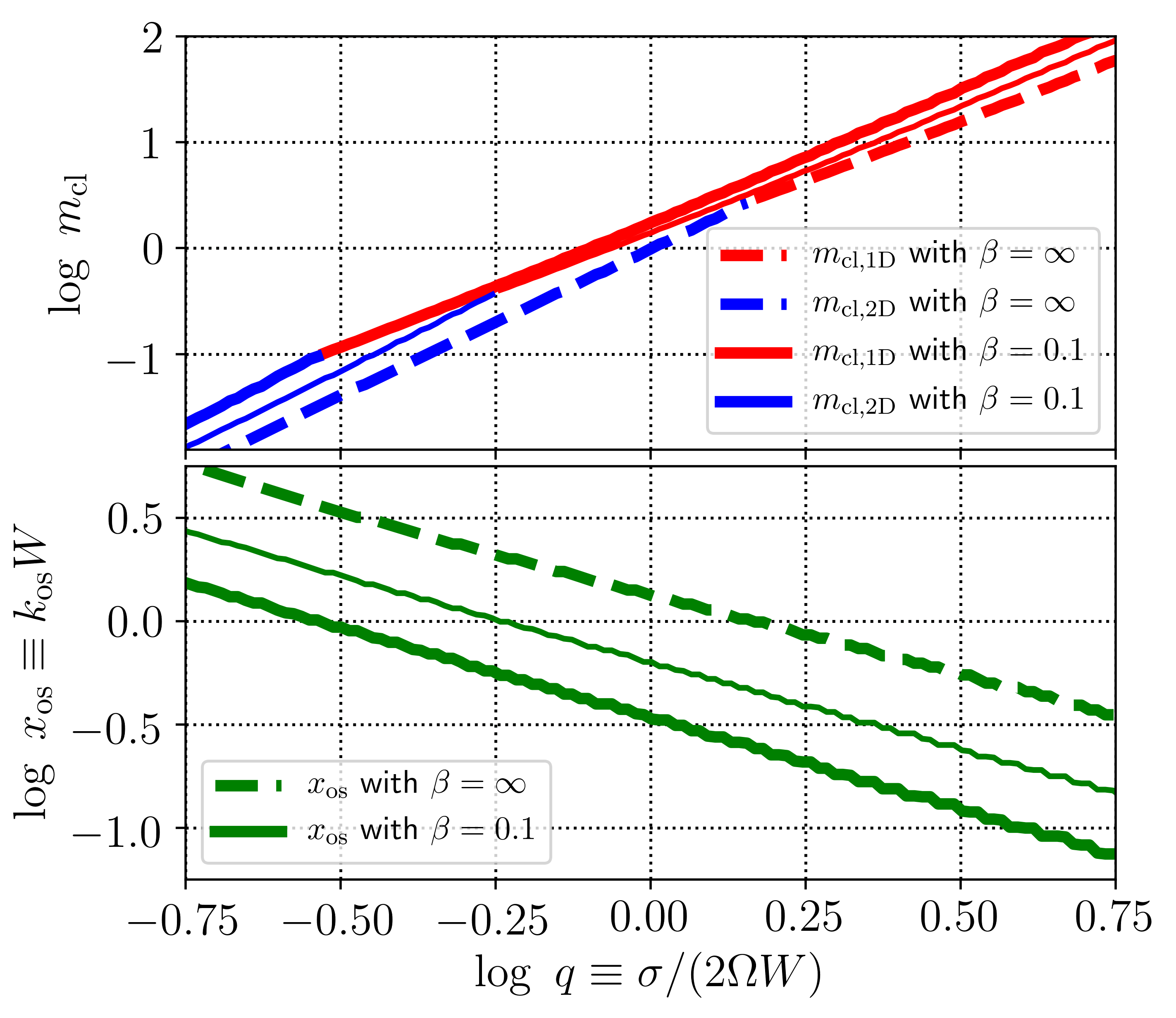}
  \caption{\textit{Top panel}: the clump mass moduli $m_{\rm cl,1D}\equiv q/(Q_{\rm sp}x_{\rm os})$ and $m_{\rm cl,2D}\equiv q/(Q_{\rm sp}x_{\rm os}^2)$ in equations (\ref{mcl1d} and \ref{mcl2d}) with $\beta=\infty$ and $0.1$; $m_{\rm cl,1D}$ and $m_{\rm cl,2D}$ are switched at $x_{\rm os}=1$ in the bottom panel. \textit{Bottom panel}: dimensionless wavenumbers $x_{\rm os}$ of the marginally unstable perturbations. In both panels, thick and thin solid lines correspond to the cases assuming the instability criterion to be $\max(s_{\rm grow})=0.1$ and $0.3$, respectively (i.e. the contours shown in Fig. \ref{DRall}).}
  \label{Mclump}
\end{figure}
In the top panel of Fig. \ref{Mclump}, we compare the clump-mass moduli $m_{\rm cl}$ as functions of $q$ between the cases of $\beta=\infty$ and $0.1$. For the magnetic case with $\beta=0.1$, we assume the practical instability criterion to be $\max(s_{\rm grow})=0.1$
(thick solid lines) and $0.3$ (thin solid lines). The values of $m_{\rm cl}$ increase only slightly even in the strong magnetic field with $\beta=0.1$. When $\beta=10$ and $1$, we find that $m_{\rm cl}$ hardly changes from the result with $\beta=\infty$. Thus, presence of toroidal magnetic field does not systematically alter masses of clumps forming via SAI if $\Omega$ and $W$ are the same. This result does not mean, however, that magnetic force is ineffective in the clump formation. The bottom panel of Fig. \ref{Mclump} shows characteristic wavenumbers $x_{\rm os}$ at the instability criteria. There, $x_{\rm os}$ is significantly small in the case of $\beta=0.1$. In the presence of strong magnetic fields, a spiral arm can be unstable at a relatively high $Q_{\rm sp}$ (i.e. low $\Sigma$ at given $\Omega$ and $\sigma$) as shown in Fig. \ref{DRall}; however the unstable wavelength $\lambda_{\rm os}$ becomes long. Thus, the large physical size of the collapsing region compensates the low density within the arm, therefore the resultant clump mass expected from the analysis is insusceptible to strength of a toroidal magnetic field. 

Using the above analysis, we estimate a typical mass of a giant clump that is expected to form by the marginal instability in our SAI model. Assuming the typical values of disc galaxies, such as $v_\phi=200~{\rm km~s^{-1}}$ and $W=0.5~{\rm kpc}$, and clump formation at $R=5~{\rm kpc}$, the factor common between equations (\ref{mcl1d} and \ref{mcl2d}) becomes $\Omega^2W^3/G=4.7\times10^7~{\rm M_\odot}$. The other factors of $8A$ in equation (\ref{mcl1d}) and  $4\pi^2$ in equation (\ref{mcl2d}) differ only by a factor of $\sim4$. Since highly turbulent gas discs of clumpy galaxies have been observed to have $\sigma\sim50~{\rm km~s^{-1}}$ \citep[e.g.][]{ggm:10,spc:11,sss:12,bgf:14,fga:17,f:17,ofg:18}, and $q\equiv\sigma/(2\Omega W)\sim1$. From the top panel of Fig. \ref{Mclump}, the clump-mass moduli are $m_{\rm cl}\sim1$ almost independent of magnetic field strength. We thus estimate the typical clump mass to be $M_{\rm cl}\sim10^7$--$10^8~{\rm M_\odot}$ from equations (\ref{mcl1d} and \ref{mcl2d}), which is approximately consistent with relatively massive clumps observed in high- and low-redshift galaxies \citep[e.g.][]{fsg:11,aob:13,ees:13,wrg:14,gfb:14,grb:18,fga:17}. Although giant clumps can significantly decrease their masses and may be disrupted after their formation, clumps with $M_{\rm cl}\sim10^8~{\rm M_\odot}$ are thought to survive even if there are feedback effects by supernovae and radiation pressure from massive stars \citep[][although see \citealt{ho:10,g:12,bmo:17}]{mdc:17}.

As shown in the bottom panel of Fig. \ref{Mclump}, wavenumber $x_{\rm os}$ of the marginally unstable perturbation decreases as $\beta$ decreases. This implies that SAI occurs at a long wavelength in a strong magnetic field and that the unstable arm tends to collapse one-dimensionally. This large-scale SAI can be expected to destructively disarrange a configuration of the arm, and such large-scale instability may tear off the arm. On the other hand, the wavelength of the marginal instability becomes short in a weak magnetic field. Such small-scale instability induced within an arm --- which is expected to collapse two-dimensionally --- might not intensely violate the spiral arm. In this case, the arm may keep its configuration although `beads on a string' structures along the arm may develop \citep[e.g.][]{ee:83,eee:18}. Thus, although strength of toroidal magnetic fields would not significantly affect a typical clump mass, it may be relevant to how violent SAI is for spiral arms.

\subsection{Other possible effects of magnetic fields}
\label{othereffects}
As we demonstrate using our simulations in Section \ref{result}, magnetic destabilisation by strong toroidal fields can lead spiral arms to fragment and form giant clumps. This result is consistent with our linear perturbation analysis presented in Section \ref{basiceq} although significant radial magnetic fields can stabilise arms in the simulations; this effect is not considered in our analysis. Our analysis assumes various simplifications, such as toroidal magnetic field without radial gradient of magnetic pressure, azimuthal perturbation (i.e. parallel to the magnetic field) propagating along spiral arm, the tight-winding approximation and rigid rotation within arm. Of course, these assumptions do not necessarily hold in simulations and real galaxies. 

Magnetic force can cause complicated effects especially when orientation of magnetic field is not parallel to perturbations. An analytic study of \citet{ko:00} has, in the context of ideal MHD, discussed various effects in rotating systems. Although their analysis does not take into account self-gravity, they consider general cases for orientations of perturbations and magnetic fields. For example, if there is significant radial gradient of magnetic pressure in a spiral arm, it exerts a force outwards even if the magnetic orientation is toroidal. As seen in our simulation results, the magnetic field strength $\beta$ varies with radius especially at inner radii although $\beta_{\rm ini}$ is uniform in our initial conditions. Because a pitch-angle of a spiral arm is not exactly zero, the arm can have significant radial gradient of $\beta$. Therefore, the magnetic pressure can push a spiral arm in the radial direction. Moreover, as we show in Section \ref{radialB}, the magnetic fields are not necessarily toroidal in spiral arms in our simulations.

If there are vertical and/or radial perturbations, gas in a spiral arm can be subject to Parker instability \citep{p:66} and/or toroidal buoyancy \citep{ko:00} which are caused by bending of magnetic field lines. It may be expected that these effects could disturb spiral arms and trigger formation of clumpy structures \citep[e.g.][]{kbp:18} although \citet{kos:02} have demonstrated, using their shearing-box MHD simulations, that Parker instability appears to play only a secondary role in disc instability. Moreover, since a galactic disc generally rotates differentially, magneto-rotational instability can also operate, which stems from radial perturbations.

Although our SAI analysis is focused on azimuthal perturbations within spiral arms, toroidal magnetic fields can also affect radial perturbations within disc regions. In the context of Toomre's instability analysis for a uniform gas disc, previous studies have presented their linear perturbation analyses taking into account toroidal magnetic fields \citep[e.g.][]{l:66,e:87,e:94,g:96,ko:01}. Considering a single-component gas disc in a toroidal magnetic field, dispersion relation for radial perturbations in a local disc region is described as
\begin{equation}
\omega^2=\left(\sigma_R^2+v_{\rm A}^2\right)k_R^2-2\pi G\Sigma_{\rm d}k_R + \kappa^2,
\end{equation}
where $\Sigma_{\rm d}$ and $\kappa$ are surface density of disc and epicyclic frequency, and $k_R$ is radial wavenumber of perturbation. Toomre's instability parameter $Q$ is accordingly modified due to the toroidal magnetic field as
\begin{equation}
Q'=\frac{\sqrt{\sigma_R^2+v_{\rm A}^2}\kappa}{\pi G\Sigma_{\rm d}}.
\end{equation}
Toroidal magnetic field thus exerts magnetic pressure in radial direction and can stabilise radial perturbations in a local disc region. If we assume that spiral arms start forming in a uniform disc, magnetic fields can suppress growth of radial perturbations and may result in formation of weak spiral arms. Because such weak spiral arms are expected to have large $Q_{\rm sp}$, in this sense, toroidal magnetic fields in the initial disc can prevent spiral arms from fragmenting and forming giant clumps. We argue, however, that magnetic fields can destabilise spiral arms \textit{after} their formation. In our simulations, the spiral arms are marginally stable when $\beta_{\rm ini}=\infty$ (see our Paper I), therefore expected to be susceptible to magnetic fields. In such cases, as our simulations show, toroidal magnetic fields can induce clump formation.

\subsection{On spiral-arm fragmentation in real galaxies and giant clump formation via SAI}
\label{reals}
Formation of giant clumps generally involves active star formation inside them. \citet{ggf:11} have estimated the total star formation rate within giant clumps in a galaxy at a redshift $z\sim2$ to be nearly fifty per cent of that within the entire galaxy.\footnote{Note that estimations of clump sizes and masses can significantly depend on observational resolutions \cite[e.g.][]{dsc:17,csr:18}. This would also be the case for estimations of star formation rates of clumps.} Thus, if these clumps are formed via spiral-arm fragmentation, magnetic fields may cause intense star formation in disc galaxies.

In \citet{pgg:17}, however, their cosmological simulations \citep[the Auriga simulations,][]{ggm:17} including ideal MHD effects have demonstrated that magnetic effects hardly change star formation histories and global evolution of their simulated galaxies. Their simulations also showed that magnetic energies within the galaxies increase with time but saturate at redshifts $z=2$--$3$, and their kinematic energies of gas significantly dominate over their magnetic energies even at $z=0$. None of the galaxies in the Auriga simulations clearly experience clumpy phases driven by disc instability, and it appears that the magnetic destabilisation is only limited in their simulations. 

The epoch of the magnetic saturation at $z=2$--$3$ shown in \citet{pgg:17}, however, coincides with the peak of abundances of clumpy galaxies \citep{sok:16}. Because galaxies are generally gas-rich at the redshifts $z=2$--$3$, magnetic effects may be influential in galactic dynamics. It has been observed that the onset of spiral galaxies in the Universe occurs at $z\simeq2$ \citep{lss:12,ee:14}. \citet{yrg:17} report their discovery of the most ancient spiral galaxy at $z=2.54$, thanks to magnification effect by gravitational lensing.\footnote{They find the spiral arms in the galaxy from a photometric image obtained from the Hubble space telescope and confirmed its disc rotation from an integral-field spectroscopy by  the Gemini North telescope.} They argue that spiral-arm structures can not be observationally resolved at this redshift without the aid of gravitational lensing and that spiral galaxies could exist at even higher redshifts. The high-redshift spiral galaxies discovered by \citet{lss:12} and \citet{yrg:17} are observed to host massive clumps within their spiral arms. Although it is uncertain whether magnetic force is significantly influential in these high-redshift spiral galaxies, we propose SAI to be a possible mechanism to trigger formation of giant clumps in these galaxies (Paper I), besides Toomre instability \citep[e.g.][]{n:98,n:99}.

\section{Conclusions and summary}
\label{conclusions}
Our study presents linear perturbation analysis for self-gravitating spiral arms in toroidal magnetic fields. Based on the SAI analysis of our Paper I, we assume ideal MHD, barotropic equation of state for gas, rigid rotation, Gaussian density distribution inside spiral arms and tight-winding approximation. Furthermore, we extend the magnetic SAI analysis to the two-component model that consists of gas and stellar components. As proposed by previous studies in the context of Toomre's instability analysis, toroidal magnetic fields can destabilise spiral arms by canceling Coriolis force and induce fragmentation of arms.

We run MHD simulations with single- and two-component disc galaxy models in isolation, and then we test our theory by adopting the simulation results to the linear analysis. We find that our analysis can characterise stable and unstable states of spiral arms. In our simulations, however, significant radial components of magnetic fields are often seen in some spiral arms. In such cases, our analysis is inapplicable, and spiral arms do not fragment and are practically stable in the simulations although our linear analysis predicts exponential growth of perturbations with finite growth time-scales. If the linear analysis ignores magnetic fields, it erroneously predicts stable states for fragmenting spiral arms. Hence, it is important to take into account the magnetic effect in order to characterise SAI more accurately. 

Using our SAI analysis, we estimate a typical mass of giant clumps forming via fragmentation of spiral arms and find that the clump mass is almost independent from strength of toroidal magnetic fields. The estimated mass is approximately $\sim10^7$--$10^8~{\rm M_\odot}$ and nearly consistent with relatively massive clumps observed in the high- and low-redshift galaxies. Hence, SAI could be a possible mechanism to form giant clumps in gas-rich galaxies, besides Toomre instability.

\section*{Acknowledgements}
We are grateful to the anonymous reviewer for his/her careful reading and fascinating comments to drastically improve the manuscript, especially for the analytic part. We thank Volker Springel for kindly providing the simulation code {\sc Arepo}. This study was supported by World Premier International Research Center Initiative (WPI), MEXT, Japan and by SPPEXA through JST CREST JPMHCR1414. SI receives the funding from KAKENHI Grant-in-Aid for Young Scientists (B), No. 17K17677. The numerical computations presented in this paper were carried out on Cray XC30 at Center for Computational Astrophysics, National Astronomical Observatory of Japan.


\appendix
\section{Physical scales of the instability}
\label{wavelength}
\begin{figure}
  \includegraphics[bb=0 0 941 785, width=\hsize]{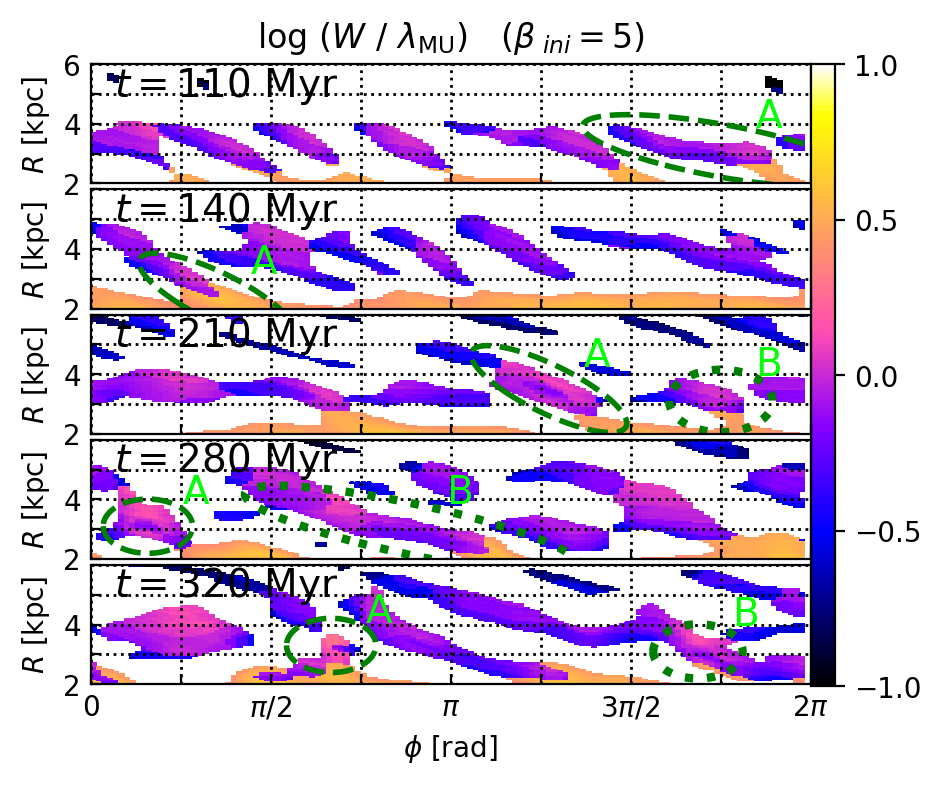}
  \caption{Ratios between arm width and wavelength of the perturbation that gives $\max(s_{\rm grow})$: $W/\lambda_{\rm MU}$ in the single-component runs with $\beta_{\rm ini}=5$.  The snapshots shown here are the same as those in Fig. \ref{Unstable}. In all spiral arms in these runs, $W/\lambda_{\rm MU}\simeq1$.}
  \label{wavelen}
\end{figure}
If a wavelength of a perturbation is significantly longer than galactocentric radius, i.e. $\lambda\gg R$, the assumptions used in our linear analysis such as ignoring curvature and pitch angle can be violated. In addition, ratio between an unstable wavelength and a width of a spiral arm determines regimes of SAI: one-dimensional or two-dimensional collapse (see Section \ref{ClumpMassDiscuss}). Therefore, it is important to look into ratios $W/\lambda_{\rm MU}$ in the fragmenting arms, where $\lambda_{\rm MU}$ is the wavelength of the most unstable perturbation that gives $\max(s_{\rm grow})$. \footnote{In this study, the most unstable wavelength is defined as $\lambda_{\rm MU}\equiv2\pi/k_{\rm MU}$, where $k_{\rm MU}$ is the wavenumber that gives $\max(s_{\rm grow})$. Hence, $x=2\pi W/\lambda_{\rm MU}$ for the most unstable perturbation.} Fig. \ref{wavelen} shows the ratios of $W/\lambda_{\rm MU}$ in the single-component run with the strong initial magnetic fields ($\beta_{\rm ini}=5$). In almost all spiral-arm regions, $W/\lambda_{\rm MU}\simeq1$, thus the most unstable wavelength is nearly comparable to the arm width. Since $W<R$, generally $\lambda_{\rm MU}\lsim R$ in the fragmenting arms. We also confirm that $W/\lambda_{\rm MU}\simeq1$ and $\lambda_{\rm MU}\lsim R$ in the other runs performed in this study.
\end{document}